\documentstyle[amsfonts,epsfig,12pt]{article}
\newcommand{\vb}[1]{{\mathbf{#1}}}
\newcommand{\r}[1]{\ref{#1}}
\newcommand{\lb}[1]{\label{#1}}
\newcommand{\bc}{\begin{center}}
\newcommand{\ec}{\end{center}}
\newcommand{\be}{\begin{equation}}
\newcommand{\ee}{\end{equation}}
\newcommand{\bea}{\begin{eqnarray}}
\newcommand{\eea}{\end{eqnarray}}
\newcommand{\ba}[1]{\begin{array}{#1}}
\newcommand{\ea}{\end{array}}
\newcommand{\bz}{{\bar{z}}}
\newcommand{\bj}{{\bar{j}}}
\newcommand{\bt}[1]{\begin{table}[ht]\centering\begin{tabular}{#1}}
\newcommand{\et}[1]{\end{tabular}\caption{\small#1}\end{table}}
\newcommand{\fig}[3]{\begin{figure}[ht]\epsfxsize=163mm\epsfbox{#1}\caption{\small #2 \label{#3}}\end{figure}}
\newcommand{\mod}{\,{\mathrm{mod}}\,}

\begin{document}

\addtolength{\baselineskip}{0.20\baselineskip}

\thispagestyle{empty}

\begin{flushright}
{\sc OUTP}-00-14P\\
hep-th/0004078\\
\end{flushright}
\vspace{.3cm}

\begin{center}

\vspace{18pt}

{\Large\bf{Toroidal Compactification in String Theory}}\\ 
\vspace{0.5 cm}
{\Large\bf{from Chern-Simons Theory}}\\

\vspace{2 truecm}

{\large  P. Castelo Ferreira\footnote{e-mail: pcastelo@thphys.ox.ac.uk}, Ian
I.\ Kogan\footnote{e-mail: kogan@thphys.ox.ac.uk}  and Bayram Tekin\footnote{e-mail: tekin@thphys.ox.ac.uk}} \\
{\it Department of Physics -- Theoretical Physics, University of Oxford\\ 1
Keble Road, Oxford OX1 3NP, U.K.} \\

\vspace{2.5 truecm}

{\sc Abstract}

\begin{minipage}{13cm}
\vspace{.5 truecm}
A detailed study of the charge spectrum of 
three dimensional Abelian Topological Massive Gauge Theory (TMGT) 
is given. When this theory is defined on a manifold with two disconnected
boundaries there are induced chiral Conformal Field Theories (CFT's)
on the boundaries which can be interpreted
as the left and right sectors of closed strings.  
We show that Narain constraints on toroidal compactification (integer, even, self-dual
momentum lattice) have a natural interpretation in purely three dimensional terms.
This is an important result which is necessary to construct toroidal compactification
and heterotic string from Topological Membrane (TM) approach to string theory.
We also derive the block structure of $c=1$ Rational Conformal Field Theory (RCFT)
from the three dimensional gauge theory.
\end{minipage}

\end{center}

\noindent

\vfill
\begin{flushleft}
\end{flushleft}
\newpage
\thispagestyle{empty}

\tableofcontents

\newpage
\setcounter{page}{1}

\section{Introduction \lb{sec.intro}}

Inherent to all known perturbative string theories are the two dimensional
conformal field theories. Moreover, keeping in mind the fact that a given string 
theory can be considered as a two dimensional sigma model (quantum field theory)
which has spacetime as its target space, it is hard to over 
emphasize the importance of $2D$ conformal field theories.
Most of the claimed beauties of string theories undoubtly reside in the fact that 
one has free and conformal theories on the worldsheet. For the closed string case,
which we exclusively work within this paper, an important feature of 
the the worldsheet theory is that left and right moving degrees of freedom decouple
from each other. One can then formulate the theory by considering two chiral conformal field theories, 
holomorphic and anti-holomorphic, living on the same two dimensional surface $\Sigma$.
The features of the non-chiral CFT, and the corresponding string theory, depend on
how one combines the holomorphic and anti-holomorphic theories. This part of the story is well 
known. In this paper we continue to develop the idea of Topological Membrane approach to
String Theory which was first set forward in
\cite{IK_0,KC_0} and worked out in many subsequent papers, where the left and right moving
sectors of closed strings live on two different surfaces ${\Sigma}_L$ and ${\Sigma}_R$. 
In other words, we are giving a finite physical thickness (be it spatial or temporal) 
to the string worldsheet which becomes now a three-manifold. 

Topological Membrane idea is based on the results of Witten~\cite{W_1}
and Moore and Seiberg~\cite{MS_1} (see also~\cite{MS_2}),
Chern-Simons theory defined on a three-manifold with a boundary is equivalent
to $2D$ conformal field theory, gauged chiral WZWN model, 
living on the boundary.
This is essentially the first example of a theory where the currently popular
holographic principle works~\cite{holog}. 
The chain of reasoning is as follows: the $3D$ bulk theory is equivalent to the boundary 
conformal theory and the later underlines a $10D$ string theory. So the 
hope is that at the very fundamental level of string theory lies a three
dimensional quantum field theory. Although it is rather speculative at this stage
this formulation might provide us with an off-shell formulation of string
theory as well as new relations between different types of strings from a more
fundamental theory.

The actual story is not as simple as it is laid above. Chern-Simons theory
in the bulk is not enough to reproduce String Theory. 
See~\cite{KC_0} for more complete details.
Besides the fact that we must induce all aspects (moduli, Liouville fields, gravitational
dressing, etc.) of  $2d$ gravity~\cite{KS_1}
there is a problem of obtaining the spectrum of toroidal compactification~\cite{N_1}
and heterotic string~\cite{GHMR_1,GHMR_2} from the original pure Chern-Simons
theory~\cite{W_1,MS_1}. This is due to the impossibility to have different right and left
charge spectrum and to induce
holomorphic CFT in one boundary without inducing anything in the other. This problem is
solved by adding a Maxwell term in the action and considering CS as the low energy
limit of TMGT~\cite{D_1}. In this way monopoles emerge in the
theory and induce new processes~\cite{Lee} which lead to local charge non conservation that
permit a right charge to have a different value than a left charge. The important point
is that bulk effects (monopoles) allow us to have winding modes in the boundary theory.
Imposing suitable boundary conditions~\cite{IK_1,LC_1} one can construct heterotic strings
in this framework.

With the  Maxwell term the bulk theory of course becomes non-topological
and the holographic principle partially works. The spectrum of the
Topological Massive Gauge Theory has a mass gap between its
excited states and the degenerate ground state. But still 
the Hilbert subspace of the ground state of Abelian TMGT is exactly equivalent 
to the Hilbert space of the full CS theory. Moreover in order to define the
quantum CS theory the Maxwell term
has to be introduced as an ultraviolet regulator, which is
set to zero after renormalization~\cite{TEKIN}. So in any case one needs the Maxwell term.

In~\cite{IK_2} the charge non conservation induced by monopoles was discussed and and these ideas were 
applied to T-duality and Mirror Symmetry in~\cite{KS_2}. Nevertheless the
problem of building the correct left-right spectrum has never been properly 
solved. The main goal of this paper is to determine the charge lattice
structure allowed by TMGT. Essentially we show that, upon imposing suitable
boundary conditions, the theory demands a charge lattice
which is exactly of the form of string theory momentum lattice.
From the non-perturbative dynamics of three dimensional
gauge theory we derive the Narain lattice spectrum.
We consider here both bosonic and heterotic string theories.

In section~\r{sec.review} we start with a short review of $3D$ TMGT and
some related topics necessary for the present work, namely some
results in Conformal Theories and String Theory. Next we move on to the main
part of this paper, section~\r{sec.bulk}, a model which
describes the dynamics of charges
propagating in the $3D$ bulk theory is built. We arrive
to some well know results in string theory but derived purely from the dynamics of
the bulk $3D$ theory. Namely the mass spectrum of toroidally compactified
closed string theory emerges.
In section~\r{sec.bc} we study a relevant issue of
gluing both $2D$ boundaries of the $3D$ manifold in order to get a
single conformal field theory in the boundary.
In section~\r{sec.cb} the underlying structure of
conformal block structure of the $c=1$ compactified bosonic RCFT and the corresponding
fusion rules are found as a result of monopole-instanton induced interactions in the bulk.
Finally in section~\r{sec.lat} the spectrum of heterotic string
and possible backgrounds are rederived in the light of these new
results.

\section{A Short Review on TMGT \lb{sec.review}}

There are several ways to derive CFT from CS theory. We choose the path integral approach
first suggested by Ogura~\cite{O_1} (see also~\cite{C_1}).
In this section we review some features of the topologically massive gauge theory defined
on a three dimensional flat manifold with a boundary.
In order to clarify some points we present some arguments derived from canonical formalism.
We present a list of the induced chiral boundary conformal 
field theories and we carefully analyze the Gauss law structure of the gauge
theory with compact gauge group $U(1)$ given by the action  
\be
S=\int_{\cal{M}} d^2z\,dt\left[-\frac{\sqrt{-g_{(3)}}}{4\gamma}F^{\mu\nu}F_{\mu\nu}+
\frac{k}{8\pi}\epsilon^{\mu\nu\lambda}A_\mu\partial_\nu A_\lambda\right]
\lb{S}
\ee
where ${\cal{M}}=\Sigma\times[0,1]$ and $\Sigma$ is a $2$ dimensional compact Euclidean manifold
with a complex structure denoted by $(z,\bz)$.
The time-like coordinate takes values in the compact domain $t\in[0,1]$. 
The indices run over $\mu=0,i$ with $i=z,\bz$. 

Under an infinitesimal variation of the fields $A\rightarrow~A+\delta A$ the action
changes by
\be
\delta S = \int_{M}\left(\frac{\sqrt{-g_{(3)}}}{\gamma}\partial_{\mu}F^{\mu\nu}+\frac{k}{4\pi}\epsilon^{\mu\lambda\nu}\partial_\mu A_\lambda\right)\delta A_{\nu}-\left[\int_\Sigma \Pi^i\delta A_i\right]^{t=1}_{t=0}
\lb{dS}
\ee
where
\be
\Pi^i= \frac{1}{\gamma}F^{0i}-\frac{k}{8\pi}\tilde{\epsilon}^{ij}A_j
\lb{P}
\ee
is the canonical momenta conjugate to $A_i$.
Note that the $2d$ antisymmetric tensor $\tilde{\epsilon}^{ij}$ is induced by the $3d$ antisymmetric tensor and
metric
\be
\tilde{\epsilon}^{ij}=\frac{-\epsilon^{0ij}}{\sqrt{-g_{(3)}}}
\ee
When referring to the usual $2d$ antisymmetric tensor we use the
notation $\epsilon^{ij}$ without the tilde.

In order the theory to have classical extrema it is necessary to impose suitable
\textit{boundary conditions} for which the second term in the variation of the action vanishes.
Let us assume that the boundary of ${\cal{M}}$ has two 
pieces, which are $\Sigma_{(t=0)}=\Sigma_L$ and $\Sigma_{(t=1)}=\Sigma_R$. 
$L$ and $R$ denote left and right.
On each of the boundaries, up to gauge transformations, one can fix 
one or both fields $A_z$ and $A_\bz$. In doing so one should add an
appropriate boundary 
action $S_B=S_{BL}+S_{BR}$ such that the new action $S+S_B$ has no boundary variation,
hence well defined classical extrema.
Note that upon canonical quantization  we have to impose
the corresponding equal time commutation relations of $\Pi$ and $A$
\be
[\Pi^i(\vb{z}),A^\bj(\vb{z'})] =g^{i\bj}\delta^{(2)}(\vb{z}-\vb{z'}) 
\lb{com_pa}
\ee
Our convention for the metric is $g^{z\bz} = g^{\bz z} = 2$.
So fixing $A^z$ we fix $\Pi^\bz$ as well and the same holds for the $A^\bz$
and $\Pi^z$ components.

On each components of $\Sigma$ the possible choices of boundary conditions
and the boundary actions are
\be
\ba{lcccc}
            & & & \\
            &\mathrm\small boundary\ conditions&\mathrm\small left\ bound.\ action&\mathrm\small right\ bound.\ action\\
            & & & \\
N.          &\delta A_z=\delta A_\bz=0&S_{BL}=0&S_{BR}=0\\
            & & & \\
C.          &\delta A_\bz=0&\displaystyle S_{BL}=-\int_{\Sigma_R}\Pi^z A_z&\displaystyle S_{BR}=\int_{\Sigma_L}\Pi^zA_z\\
            & & & \\
\bar{C}.\ \ &\delta A_z=0&\displaystyle S_{BL}=-\int_{\Sigma_R}\Pi^\bz A_\bz&\displaystyle S_{BR}=\int_{\Sigma_L}\Pi^\bz A_\bz\\
            &\, &\, & \,   
\ea
\lb{bc}
\ee
There are  nine allowed choices:
$NN$, $NC$, $CC$, $C\bar{C}$, and so on. The first letter denotes the type of left
boundary conditions and the second one the right type.
The $N$ boundary condition stands for \textit{Non-Conformal} or \textit{Non-Dynamical},
$C$ stands for \textit{Conformal} and $\bar{C}$ for \textit{anti-Conformal}
and are related to the kind of CFT we obtain in the boundaries when we choose them.
Note the importance of the $F^2$ term, it gives the theory four independent canonical coordinates,
opposed to the two of the pure Chern-Simons theory (where one of the $A$'s is canonically
conjugate to the other). This fact allow us to fix one of the $A$'s and corresponding
$\Pi$ without fixing the other two allowing us to have different boundary conditions
in opposite boundaries.
For further details we refer the reader to~\cite{IK_1,LC_1}. This topic will be addressed again
in section~\r{sec.bc}.

By Faddeev-Popov procedure we can fix the gauge $A_\mu=\bar{A}_\mu+\partial_\mu\Lambda$.
The action and path integral factorizes as
\be
Z=\int{\mathcal{D}}\bar{\vb{A}}\Delta_{FP}\delta\left(F(\bar{\vb{A}})\right)e^{i(\bar{S}+\bar{S}_B)[\bar{\vb{A}}]}\int{\mathcal{D}}\lambda{\mathcal{D}}\chi e^{i\delta\bar{S}_L[\chi]+i\delta\bar{S}_R[\lambda]}
\ee
where in the boundaries the gauge parameters $\Lambda(t=0)=\chi$ and $\Lambda(t=1)=\lambda$
become dynamical degrees of freedom which decouple from the bulk theory.

Now let us find the allowed charges in this theory. The compactness 
of the gauge group has to be taken into account and it turns out to be of fundamental importance.  
Defining the electric and magnetic fields as
\be
\ba{rcl}
E^i&=&\displaystyle 
\frac{1}{\gamma}F^{0i}=\Pi^i+\frac{k}{8\pi}\tilde{\epsilon}^{ij}A_j\vspace{.2 cm}\\
B&=&\displaystyle\epsilon^{ij}\partial_iA_{j}
\ea
\lb{EB}
\ee
(For the sake of notational simplicity we have omitted the complex conjugation 
on the indices)
The  commutation relations read
\be
\ba{rcl}
\left[E^i(\vb{z}),E^j(\vb{z'})\right]&=&-i{k\over{4\pi}}\epsilon^{ij}\delta^{(2)}(\vb{z}-\vb{z'})\vspace{.2 cm}\\
\left[E^i(\vb{z}),B(\vb{z'})\right]&=&-i\epsilon^{ij}\partial_j\delta^{(2)}(\vb{z}-\vb{z'})
\ea
\lb{com_EB}
\ee
If there is an external charge the Gauss law is
\be
\partial_i E^i+{k\over{4\pi}}B=\rho_0
\lb{gauss}
\ee
In the quantum theory this equation needs to be satisfied by the physical
states. So following~\cite{IK_2} we can define the generator of time independent
gauge transformations $U$
\be
U=\exp\left\{i \int_\Sigma\Lambda(\vb{z})\left(\partial_i E^i+{k\over{4\pi}} B-\rho_0\right)\right\}
\lb{U}
\ee
$\Sigma$ stands for a generic fixed time slice of ${\cal{M}}$.
Since The gauge group is compact $\Lambda$ is identified with an
angle in the complex plane such that
\be
\ba{rcl}
\ln(z)&=&\ln|z|+i(\Lambda(z)+2\pi n)\vspace{.2 cm}\\
\partial_i\Lambda(z)&=&-\epsilon_{ij}\partial_j\ln|z|
\ea
\ee
where the second equation follows from the Cauchy-Riemann equations. This last condition
on $\Lambda$ will restrict the  physical Hilbert space in the compact
theory~\cite{AK_1}. Let us define a new operator
\be
V(\vb{z_0})=\exp\left\{-i\int_\Sigma d^2z\left[\left(E^i+\frac{k}{4\pi}\epsilon^{ij}A_j\right)\epsilon^{ik}\partial_k\ln|\vb{z_0}-\vb{z}|-\Lambda(\vb{z_0}-\vb{z})\rho_0\right]\right\}
\lb{V}
\ee
The physical states of the theory must be gauge invariant (under $U$
operator) as well eigenstates of this new local operator.
Using the identity $\partial_k\partial_k\ln|z|=2\pi\delta(z)$ and the
commutation relations (\r{com_EB}) for $E$ and $B$ we obtain 
\be
\left[B(\vb{z}),V^n(\vb{z_0})\right]=2\pi nV^n(\vb{z_0})\delta^{(2)}(\vb{z}-\vb{z'})
\ee
This means that the operator $V$ creates a pointlike magnetic vortex at $\vb{z_0}$
with magnetic flux
\be
\int_\Sigma B=2\pi n\ \ \ \ \ n\in\mathbb{Z}
\lb{flux}
\ee

As stated before the $F^2$ term is fundamental for the existence of these tunneling processes
that hold local charge non conservation, see~\cite{Lee,AK_1} for further details.
Instantons in three dimensions are the monopoles in four
dimensions. So in the rest of the paper  
we will make use of the terms instanton and monopole without any distinction.

Using the functional Schr\"{o}dinger representation $\Pi^i=i\delta/\delta
A_i$ and imposing the condition that phase acquired by a physical state under
a gauge transformation
be single valued we obtain the charge spectrum
\be
q=m+\frac{k}{8\pi}\int_\Sigma B=m+\frac{k}{4}\,n \ \ \ \ \ m,n\in\mathbb{Z}  
\lb{q}
\ee
As it can be seen from the above formula for a generic, non-integer, value of $k$ 
the allowed charges are non-integers. In principle in the compact Abelian
theory one expects that the charges are quantized as it happens in the
Maxwell theory. But the existence of the Chern-Simons term changes this
picture, the charges are quantized trough $m$ and $n$ dependence
but we consider $k$ itself not quantized. As will be explained below this fact
is only compatible (and demands) with the existence of
a non compact gauge sector in the theory.

A charge $q$ propagating in the bulk can interact with one monopole with flux (\r{flux})
changing by an amount
\be
\Delta q =\frac{k}{2}\, n
\lb{dq}
\ee

The path of a charge in the bulk can be thought as a Wilson line.    
The phase induced by the linking of two Wilson lines carrying charges
$q_1$ and $q_2$ is
\be
<W_{q_1}W_{q_2}>=\exp\left\{2\pi i \frac{2}{k}\, q_1\, q_2\, l\right\}
\lb{df}
\ee
where $l$ is the linking number between the lines. The above computation
was done in the limit of vanishing Maxwell term and with the assumption that
there is no self-linking for the individual Wilson lines.
The connection between the boundary CFT's and the bulk theory can be
achieved by noting that the bulk gauge fields become pure gauges in the
boundary and the Wilson lines in the boundary are none other than the 
vertex operators in $\Sigma_L$ and $\Sigma_R$ with momenta $q$
\be
\ba{c}
V_{R,q_R}=\left.\exp\left\{-iq\int A_\nu dx^\nu\right\}\right|_{\Sigma_R}=\exp\left\{-iq_R\lambda\right\}\vspace{.2 cm}\\
V_{L,q_L}=\left.\exp\left\{-iq\int A_\nu dx^\nu\right\}\right|_{\Sigma_L}=\exp\left\{-iq_L\chi\right\}
\ea
\lb{vertex}
\ee
The conformal dimensions of these vertices is
\be
\ba{c}
\Delta_R=\frac{q_R^2}{k}\vspace{.2 cm}\\
\Delta_L=\frac{q_L^2}{k}
\ea
\lb{confd}
\ee
$k$ appears because the induced action in the boundary is that
of a chiral string action multiplied by $k$. 
To see the Wilson line and vertex operator correspondence let us 
1obtain the above  result from the bulk theory.
Consider two Wilson lines carrying charges $q$ and $-q$
propagating from one boundary to the other corresponding to two
vertex insertion in the boundary with momenta $q$ and $-q$, see figure 
(\r{fig0}).
\fig{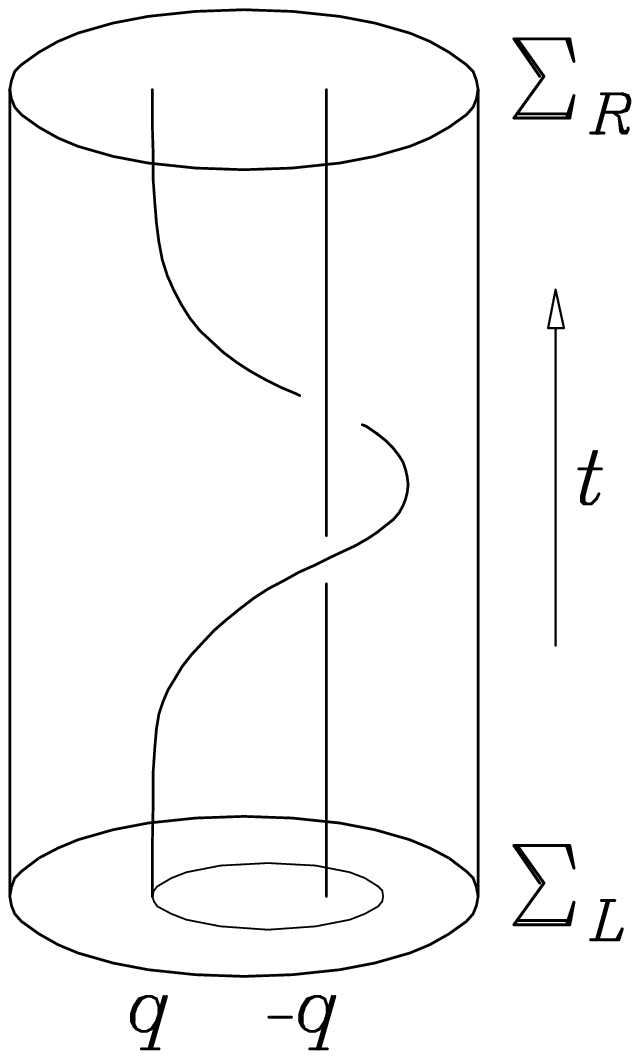}{Two charges propagating through the bulk are represented as two Wilson lines,
By a rotation of one charge around the other we induce one linking ($l=1$) in the bulk.}{fig0}

Two point correlation of these two vertices follows as usual
\be
\ba{c}
\displaystyle<V_{L,q}(\vb{z_1})V_{L,-q}(\vb{z_2}))>=\frac{1}{{\vb{z_{12}}^{2\Delta}}}\vspace{.1 cm}\vspace{.2 cm}\\
\displaystyle <V_{R,q}(\vb{\bz_1})V_{R,-q}(\vb{\bz_2}))>=\frac{1}{{\vb{\bz_{12}}^{2\bar{\Delta}}}}\\
\ea
\lb{OPE2}
\ee
A rotation of one charge (vertex) around the other in one boundary
induces a phase of $-4\pi i \Delta$ in (\ref{OPE2}).  
In the bulk this rotation induces one linking of the Wilson
lines ($l\rightarrow l+1$), so (\r{df}) gives an Aharonov-Bohm phase change of $-4\pi i q^2/k$.
Identifying these two phases we conclude that (\r{confd}) follows. 
Let us recall that in terms of the three dimensional theory Polyakov~\cite{P_1} pointed out 
that $\Delta$ is the transmuted spin of a charged particle which exists because of the
interaction with Chern-Simons term. So we identify the conformal dimension of the 
boundary fields with the transmuted spin of the bulk charges.  
For later use let us generalize this result two three charges $q_1$, $q_2$ and
$q_3=-q_1-q_2$. We will always take the net charge to be zero. 
Three point correlation of three vertices reads 
\be
<V_{q_1}(\vb{z_1})V_{q_2}(\vb{z_2})V_{q_3}(\vb{z_3}))>=\frac{1}{{\vb{z_{12}}^{\Delta_1+\Delta_2-\Delta_3}{\vb{z_{13}}^{\Delta_1+\Delta_3-\Delta_2}}}{\vb{z_{23}}^{\Delta_2+\Delta_3-\Delta_1}}}
\lb{OPE3}
\ee
where the charge-conformal dimension identifications are  
\be
\ba{rcl}
\Delta_1+\Delta_2-\Delta_3&=&-\frac{2}{k}q_1q_2\vspace{.2 cm}\\
\Delta_1+\Delta_3-\Delta_2&=&\frac{2}{k}q_1(q_1+q_2)\vspace{.2 cm}\\
\Delta_2+\Delta_3-\Delta_1&=&\frac{2}{k}q_2(q_1+q_2)
\ea
\lb{confd3}
\ee

The anti-holomorphic three point correlation follows trivially.
It is clear that if any one of the vertices is rotated  around one of the other two; the phase factor
induced in the three-point function will be the same as  (\r{df})
which comes from the linking of Wilson lines in the bulk.
From the point of view of CFT the sum of conformal factors
must be integer in order to have single valued OPE's. This result is
going to be derived from the bulk theory in the next section independent
of the boundary. By a simple argument let us show that we can only have integer
conformal dimensions for each vertex.
At the level of the  two point correlation
functions we can have  two integer $\Delta$'s  or two half integer $\Delta$'s. 
For three point functions it is necessary to have  three integer $\Delta$' or two
integer and one half integer $\Delta$'s. One can then consider the four point function
and decompose it to  two pairs which are well separated from each other. In this case 
the conformal dimensions can be integers and half-integers. Suppose that at least two
of the  vertices have  half-integer conformal dimensions.
Then one vertex from a pair can be 
adiabatically moved to the other pair and a three point function with a total 
half-integer conformal dimension will be formed. But this is not permitted as argued above.
For this reason we exclude the
existence of vertices with half integer conformal dimension.
The same arguments follow for the charges from the point of view of the $3d$~bulk theory.

Let us remember that we have two independent chiral conformal field theories on two different
surfaces up to this point. In order to get a  non-chiral CFT (as in string theory)
we have to identify in  some way the two boundaries. The most obvious way to define
non-chiral vertex operators is  
\be
V_{q_R,q_L}(\vb{z},\vb{\bz})=V_{R,q_R}(\vb{z})V_{L,q_L}(\vb{\bz})
\ee
with the conformal dimension
\be
\Delta_R+\Delta_L=\frac{1}{k}(q_R^2+q_L^2)
\lb{confdLR}
\ee
Note that we have a little bit of difference in the nomenclature.  In most of the literature 
$\Delta_R$ and $\Delta_L$ are called conformal weights and $\Delta_R-\Delta_L$ is called the
spin and the sum~(\r{confdLR}) is called the conformal dimension. Here we call $\Delta_R$ and $\Delta_L$
conformal dimensions as well when we refer to the chiral vertices.
From now on we will change our notation to $\Delta=\Delta_R$ and $\bar{\Delta}=\Delta_L$,
the same for the charges. 

For generic $k$ and $q$'s, $\Delta$ in (\r{confd}) and 
the sums in (\r{confd3}) are not integers. This is not a problem,
actually is very welcome since it is simply the statement that we need
something else in our theory, a new non
compact gauge group sector. Or in terms of string
theory, non compactified dimensions. So
in addition to the action (\ref{S}) with compact gauge group $U(1)$ we will consider the
following action with non compact gauge group $U(1)^D$
\be
S_{D}=\int_M d^2z\,dt\left[-\frac{1}{4\gamma'}f^{\mu\nu}_Mf_{\mu\nu}^M+\frac{k'\delta_{MN}}{8\pi}\epsilon^{\mu\nu\lambda}a^M_\mu\partial_\nu a^N_\lambda\right]
\lb{SD}
\ee
where $N$ and $M$ run from $0$ to $D-1$. We could take generally a coupling tensor $K'_{MN}$,
for the purposes of this work is enough to consider the above $K'_{MN}=k'\delta_{MN}$.
We will consider the actions (\ref{S}) and (\ref{SD}) together. 
A generic (meaning that not necessarily the tachyon operators we considered before) non-chiral 
vertex of the boundary CFT is of the following form
\be
\partial_z^s\zeta\partial_\bz^{s'}\zeta\left(\prod_i\partial_z^{r_i}\Omega^M\right)\left(\prod_i\partial_\bz^{r'_i}\Omega^M\right)\exp\left\{-i\left[(q+\bar{q})\zeta+p_M\Omega^M\right]\right\}
\lb{VD}
\ee
The fields $\zeta$ and $\Omega^M$ correspond to the gauge parameters of $A$ and $a^M$ respectively.
Levels of the vertex operators are integers defined as
\be
\ba{rcl}
L      &=&s+\sum_i r_i\vspace{.2 cm}\\
\bar{L}&=&s'+\sum_i r'_i
\ea
\ee

The exponential part may be represented as the bulk Wilson
line propagating from boundary to boundary
\be
W_{D}=\exp\left\{-i\int\left[qA_\nu+p_Ma^M_\nu\right]dx^\nu\right\}
\lb{WD}
\ee
The mass shell condition for the boundary vertex is
\be
p_M p^M=- \mbox{mass}^2
\lb{mass}
\ee
The mass spectrum in CFT's is built out of the allowed values for the conformal dimension
of the fields or the operators. In particular the vertex operators (\r{VD}) have conformal
dimensions
\be
\ba{rcl}
\Delta&=&\displaystyle \frac{q^2}{k}+\frac{p^2}{k'}+L\vspace{.2 cm}\\
\bar{\Delta}&=&\displaystyle\frac{\bar{q}^2}{k}+\frac{p^2}{k'}+\bar{L}
\ea
\lb{weights}
\ee
Due to conformal invariance  one has $\Delta=\bar{\Delta}=1$.
To get the usual String theory normalization one should replace $k'=4/\alpha=k/R^2$
and take the average of both equations (\r{weights})
\be
\mbox{mass}^2=-p^2=\frac{(q^2+\bar{q}^2)k'}{2k}+\frac{k'}{2}(L+\bar{L}-2)=\frac{m^2}{R^2}+\frac{R^2n^2}{\alpha^2}+\frac{2}{\alpha}(L+\bar{L}-2)
\lb{mshell}
\ee
Subtracting one equation from the other in (\r{weights})  one gets
\be
\frac{q^2-\bar{q}^2}{k}+L-\bar{L}=nm+L-\bar{L}=0
\lb{spin}
\ee
Let us call (\r{mshell}) and (\r{spin}) the mass shell condition and spin condition
respectively. For further details see \cite{POL_1}.
We used the explicit forms of the charges corresponding to
momentum spectrum of string theory, in our normalization
\be
\ba{c}
q=m+\frac{k}{4}n\vspace{.2 cm}\\
\bar{q}=m-\frac{k}{4}n
\ea
\ee
But remember that the allowed charges in the theory are of the $q$-form.
In principle $\bar{q}$ is not related to $q$ and it should be of the form
$\bar{q}=m'+n'k/4$. As it will be explained in detail in the rest of the paper we have 
the following picture in mind. A charge $q$ is inserted in one of the boundaries and
it goes through the bulk interacting with the gauge fields
until it reaches the other boundary. During its journey through the bulk its
interactions with (in)finite many monopole-instantons induce a change of its charge 
$\Delta q=Nk/2$. The the charge emerges in the other boundary as $\bar{q}=q+Nk/2=m+(n+2N)k/4$.
If $N=-n$ we have $\bar{q}=m-nk/4$, this is what we are looking for.
But why N must be $-n$? How do the monopoles know a priori they have to interact
by this amount with the initial charge $q$?
In principle it could exist any other process holding $N\neq-n$ that would generate
some charge $\bar{q}$ leading us in disaster. We would get the pair $(q,\bar{q})$
with no correspondence with any physical sensible momentum pair $(P_R,P_L)$ of
string theory. So we have to show that the full $3d$-amplitudes $q\rightarrow\bar{q}$
in the theory corresponding to unwanted processes $\bar{q}\neq m-nk/4$ are vanishing!
This crucial property shows that seemingly independent chiral field theories on
different boundaries are actually related when the non-perturbative excitations in the bulk
are taken into account. This fact is closely related to considering the Maxwell Chern-Simons
theory instead of the pure Chern-Simons and the presence in the bulk of monopole
processes. In the next section we shall prove that the theory indeed has this property.

\section{Propagation in the bulk \lb{sec.bulk}}

Our aim in this section is to investigate the changes that a charge undergo when it travels
from one boundary to the other. As it will be shown the bulk theory imposes ,independent of
what the boundary conditions are, restrictions for the charges.  
We are assuming irrational $k$ in this section and in section~\r{sec.cb} the case of 
rational $k$ will be addressed.

We are dealing with a closed manifold ${\cal{M}}$ with zero total charge on it.
The instantons can induce a charge non-conservation locally only. 
So let us insert a pair of charges $Q_1=-Q_2$ in one
boundary, say $\Sigma_L$. They will travel through the bulk and emerge in the opposite boundary,
$\Sigma_R$ as two new charges $\bar{Q}_1=-\bar{Q}_2$. In their paths the charges can interact
with the bulk fields. Their paths can link but we assume 
there is no self-linking and the charges do not interact with each other. 

Translating this to a more formal language. We have two open Wilson lines, each of them
have their ends attached to different boundaries. The Wilson line represents the path of a charged
particle. Leaving aside the perturbative interactions in the bulk, there are charge non-conserving
interactions with the monopole-instantons. We will represent these interactions as insertions of instantons
along the Wilson lines. Each instanton on the path of the particle changes the sign of the particle as mentioned 
before. So the physical
picture is that we have a Wilson line on which the charge varies from point to point. But this variation is not random
and it is induced by instantons. One can in principle consider chains of Wilson lines which is essentially the 
same picture. In between two insertions the two lines (minimal Wilson line segments with constant charges on them)
can link to each other changing the correlator  of the lines according to~(\r{df}). So we assume that we can
separate the insertions of instantons from the linkings of the lines. Or an other way to put is that insertions 
of instantons are done at the ends of the Wilson line segments. 

\fig{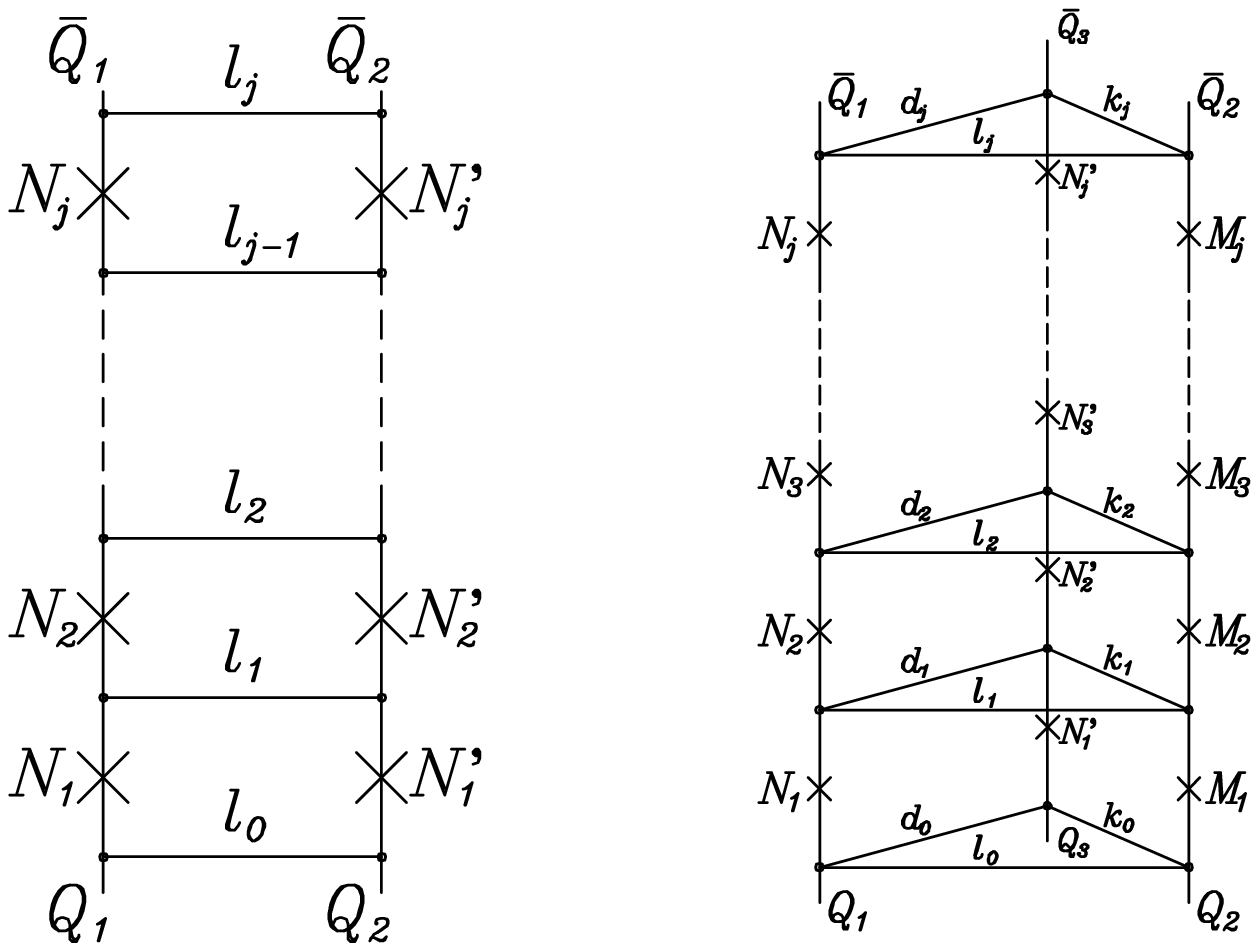}{Two and three charges propagation through the bulk. The $N$'s, $M$'s,
and $N'$'s represent the flux of instantons and the $l$'s, $d$'s and
$k$'s the linking number of the Wilson lines between insertions.}{fig1}

A generic propagation of two such charges with $j$ instanton
interactions and $j+1$ sets of linkings in the bulk is represented as a planar
diagram in figure~\r{fig1}. The insertions are represented by crosses and the linkings  
by horizontal lines. The full physical picture is obtained by taking the limit
$j\rightarrow\infty$ which allows infinitely many instanton insertions on a line.
Since we assume global charge conservation, $\bar{Q}_1-Q_1=Q_2-\bar{Q}_2$
and  the upper charges are the result of all
the monopole contributions $\bar{Q}_1=Q_1+\sum k/2 N_i$. So we have to impose the condition
\be
\sum_{i=1}^{j}N_i=N =-\sum_{i=1}^{j}N'_i
\lb{N}
\ee
Each pair of charges can be formally interpreted as a quantum state $(Q_1,-Q_1)$ and its propagation
from one boundary to the other boundary can be considered as a time evolution to the state $(\bar{Q}_1,-\bar{Q}_2)$. 
Then for each initial $Q_1$ and final $\bar{Q}_1$ we can built a transition coefficient
$C[Q_1,N]$ which is to be interpreted as the square root of probability.
\be
(Q_1,-Q_1)\rightarrow C[Q_1,N]\,(Q_1+\frac{k}{2} N,-Q_1-\frac{k}{2} N)
\ee
where we already make use of condition~(\r{N}).
$C$ is a function of $Q_1$ and $N$ only but in order to find it we need to take into account
all the possible processes in the bulk, monopole interactions and linkings.
To be able to extract some information about the boundary let us insert
a new variable $l$, which denotes the total linking number.  
Up to normalization factors $C$ can be written as 
\be
C[Q_1,N]=\sum_{\tilde{l}=-\infty}^\infty\int dl\, \delta(\tilde{l}-l)\,c[Q_1,N,l]
\lb{C}
\ee
with the definition
\be
l=l_0+\sum_{i=1}^jl_i
\lb{l}
\ee

It is now about time to find how the infinitely  many bulk processes contribute to this coefficient.
We know that interacting with the monopoles labeled by 
$i$ the pair of charges will go through the following transition  $(q,q')\rightarrow(q+k/2 N_i,q'+k/2 N'_i)$ according
to (\r{dq}). Between each monopole insertions the linking will induce a phase change
of $4\pi i qq'l_i/k$ according to (\r{df}). The total change of the charges is already taken in
account through $N$ dependence on $C$. But the processes that we are summing over must
be selected in order to fulfill the requirement that the total monopole
contribution is $N$ and that the total linking number is $l$. 
The coefficient $c$ is then the sum of the phases
\be
c[Q_1,N,l]=\sum_{N_a,N'_b}\sum_{l_i,l_0}\exp\{2\pi i \Phi[N_a,N'_b,l_i,l_0]\}
\lb{c}
\ee
Where the sums over $N_a$ and $N'_b$ stand for all the allowed configurations obeying (\r{N})
and the sum over $l_i$ and $l_0$ for the ones obeying (\r{l}).
The phase, $\Phi$ , for each process is simply the sum of the several phases induced at each step $i$
(see figure~\r{fig1}) along the propagation in the bulk
\be
\Phi=\frac{2}{k}\left\{-Q_1^2l_0+\sum_{i=1}^{j}\left(Q_1+\frac{k}{2}\sum_{b=1}^iN_b\right)\left(-Q_1+\frac{k}{2}\sum_{a=1}^iN'_a\right)l_i \right\}
\ee 
Writing the last term of the sum in $i$ using (\r{N}), i.e. for $i=j$, and expanding the
products we obtain
\be
\ba{rcl}
\Phi&=&\displaystyle
    \frac{2}{k}\left\{-Q_1^2\,l_0-Q_1^2\sum_{i=1}^{j}\,l_i+\left(Q^2_1-\left(Q_1+\frac{k}{2}N\right)^2\right)l_j\right. \vspace{.2 cm}\\
&+&\displaystyle
   \left.\frac{k}{2}\sum_{i=1}^{j-1}\left(Q_1\sum_{a=1}^{i}(N'_a-N_a)+\frac{k}{2}\sum_{a=1}^iN'_a\sum_{b=1}^iN_b\right)l_i\right\}
\ea
\lb{f}
\ee
Note that the phase is no longer $N_j$ dependent, we replaced it by $N$ dependence.
We further need to get the $l$ dependence of $c$. In the following
discussion we are going to investigate the cases for which the coefficient is non-vanishing.
Bearing in mind that we are dealing
with formal series (which are divergent in general) different ways of factorizing the sums is safer. So we will
depict three different factorizations below,
$c$ will factorize into a $l$ dependent phase and
an independent phase factor.  Unless otherwise stated from now on
the index $i$ will run from $1$ to $j-1$. 

The most obvious way to factorize the sum is to eliminate $l_0=l-\sum l_i-l_j$. Then (\r{c}) factorizes as 
\be
\ba{rcl}
c[Q_1,N,l]&=&\displaystyle\exp\left\{-2\pi i \frac{2}{k} Q_1^2\,l\right\}\times\vspace{.2 cm}\\
&\times&\displaystyle\sum_{N_{a}=-\infty}^{\infty}\sum_{N'_{b}=-\infty}^{\infty}\prod_{i=1}^{j-1}\sum_{l_{i}=-\infty}^\infty\sum_{l_j=-\infty}^\infty\exp\{2\pi i \Phi'_0[N_a,N_b,l_i,l_j]\}
\ea
\lb{c0}
\ee
The indices $a$ and $b$ run from $1$ to $j-1$.
Since the phase $\Phi'_0$ is the sum of several phase changes $\phi$ we can rewrite the second
factor as
\be
\sum_{N_{a},N'_{b}}\left(\prod_{i=1}^{j-1}\sum_{l_{i}=-\infty}^\infty\exp\{2\pi i\phi_i\,l_i\}\right)\left(\sum_{l_j=-\infty}^\infty\exp\{2\pi i \phi_j\,l_j\}\right)
\ee
where
\be
\ba{rcl}
\phi_i&=&\displaystyle
      Q_1\sum_{a=1}^{i}(N'_a-N_a)+\frac{k}{2}\sum_{a=1}^iN'_a\sum_{b=1}^iN_b\vspace{.2 cm}\\
\phi_j&=&\displaystyle
      \frac{2}{k}\left\{Q_1^2-\left(Q_1+\frac{k}{2}N\right)^2\right\}
\ea
\lb{f0}
\ee
Now we can consider each of the sums over $l_j$ and the $l_i$'s independently.
If one of them is zero  $c$ will be vanishing. Note that we didn't worry too much about normalization
factors but each of these phase sums must be normalized to a number between $0$ and $1$ in order that one has
the interpretation of $C^2$ as a probability. 
Further, if we take the limit of $j\rightarrow\infty$ these sums will
become integrals. We can investigate if they are zero or not by using the identity
\be
\sum_{q=-\infty}^\infty\exp\{2 \pi i\, \phi\, q\}=\sum_{p=-\infty}^\infty\delta(\phi-p)
\lb{fourier}
\ee
The sum over delta functions is zero if
$\phi$ is not an integer. The conclusion is then that, for every $i$,
$\phi_i$ must be an integer.
Summing over $l_j$ we obtain the restriction
\be
\frac{2}{k}\left\{Q_1^2-\left(Q_1+\frac{k}{2}N\right)^2\right\}\in \mathbb{Z}
\lb{int0}
\ee
Replacing $Q_1$ by its form (\r{q}), expanding and getting rid of the even integer term $2mN$ which does not change
anything, the former condition reads
\be
\frac{k}{2}N\left(n+N\right)\in \mathbb{Z}
\lb{int01}
\ee
For irrational $k$ the only two solutions are
\bea
N&=&0\lb{nN00}\vspace{.2 cm}\\
N&=&-n\lb{nN0n}
\eea
The physical meaning of these results will be given a little later.
The remaining conditions in the $\phi_i$'s will allow us to build the
intermediate processes, these will be addressed in the end of this section.

Now an other way to carry out the sums is to
eliminate $l_j=l-l_0-\sum l_i$. Then  (\r{c}) factorizes as
\be
\ba{rcl}
c[Q_1,N,l]&=&\displaystyle\exp\left\{-2 \pi i \frac{2}{k}\left(Q_1+\frac{k}{2}N\right)^2l\right\}\times\vspace{.2 cm}\\
&\times&\displaystyle\sum_{N_{a}=-\infty}^{\infty}\sum_{N'_{b}=-\infty}^{\infty}\prod_{i=1}^{j-1}\sum_{l_{i}=-\infty}^\infty\sum_{l_0=-\infty}^\infty\exp\{2\pi i \Phi'_j[N_a,N_b,l_i,l_0]\}
\ea
\lb{cj}
\ee
Performing the sum over $l_0$ we obtain the same conditions (\r{int0}) and (\r{int01})
on $N$.
There are extra $2N$ and $N^2$ factors but they do not change the coefficient $c$.
We can check this explicitly by replacing $Q_1$ by its form (\r{q}).
Apart from an irrelevant factor of 
$4\pi i m N$ one gets the same result as for~${\Phi'}_0$. 

An other way of factorization is as the following. Write (\r{f}) in terms of $l^\pm$
\be
l_0^+ = l_0+l_j  \hskip 1.5cm l_0^- = l_0-l_j
\lb{lpm}
\ee
and take into account that $l_0^+=l-\sum l_i$.
Then (\r{c}) factorizes as
\be
\ba{rcl}
c[Q_1,N,l]&=&\displaystyle\exp\left\{-2 \pi i \frac{2}{k}\left(Q_1^2+\left(Q_1+\frac{k}{2}N\right)^2\right)\frac{l}{2}\right\}\times\vspace{.2 cm}\\
&\times&\displaystyle\sum_{N_{a}=-\infty}^{\infty}\sum_{N'_{b}=-\infty}^{\infty}\sum_{l_{i}=-\infty}^\infty\sum_{l_0^-=-\infty}^\infty\exp\{2\pi i \Phi'_+[N_a,N_b,l_i,l_0^-]\}
\ea
\lb{c+}
\ee
Performing the sum over $l_0^-$ we obtain a different condition
\be
\frac{2}{k}\left\{Q_1^2-\left(Q_1+\frac{k}{2}N\right)^2\right\}\in 2\mathbb{Z}
\lb{even}
\ee
Expanding $Q_1$ this gets to
\be
\frac{k}{2}N\left(n+N\right)\in 2\mathbb{Z}
\lb{int0+}
\ee
For generic $k\in\mathbb{R}$ we obtain the same solutions as (\r{int0}) and
apparently this condition turns out to hold the same results of (\r{int01}). It
will become clear that this last result is different from
the previous two cases.
As before the extra $N$ and $N^2/2$ factors do not change the coefficient.

It is now time to explain the physical implications of the three previous results.
The phase factors in front of $l$ are
none other than the conformal dimensions~(\r{confd}) of the vertex operators
inserted in the boundary CFT's. When we eliminated the sum over $l_0$
we obtained~(\r{c0}). Extracting the $l$ phase factor we get  
\be
\frac{2}{k}Q_1^2=2\Delta=\Delta_1+\Delta_2
\lb{confd0}
\ee
Upon the elimination of $l_j$ sum we obtained the factorization~(\r{cj}). 
The phase factor reads
\be
\frac{2}{k}\bar{Q}_1^2=2\bar{\Delta}=\bar{\Delta}_1+\bar{\Delta}_2
\lb{confdj}
\ee
And for the (\r{c+}) factorization where we eliminated $l_0^+$ sum, we
obtain 
\be
\frac{1}{k}(Q_1^2+\bar{Q}_1^2)=\Delta+\bar{\Delta}
\lb{confd+}
\ee

Note that we cannot extract any other factors, If we try to eliminate
any other combination of $l_0$ and $l_j$ we would get $c=0$. Eliminating
some combination of $l_i$'s we will always end up extracting one of
the above factors or get again $c=0$. The solutions for the monopole
interactions (\r{ab}) and (\r{bb}) computed below assure this result.

In order $C$ not to be zero it is necessary to get integer conformal
dimensions as stated in section~\r{sec.review}. Consider then adding extra non compact gauge
fields. One Wilson line (\r{WD}) carries now charges from all the $U(1)$'s.
Summing over $l$ for the three previous cases we obtain the
conditions
\be
\ba{cccc}
r&=&\displaystyle 2\left(\Delta+\frac{p^2}{k'}\right)&\in\mathbb{Z}\vspace{.2 cm}\\
s&=&\displaystyle 2\left(\bar{\Delta}+\frac{p^2}{k'}\right)&\in\mathbb{Z}\vspace{.2 cm}\\
t&=&\displaystyle \Delta+\bar{\Delta}+\frac{2p^2}{k'}&\in \mathbb{Z}
\ea
\lb{rst}
\ee
Subtracting the second equation from the first we obtain
\be
r-s=2(\Delta-\bar{\Delta})=\frac{2}{k}(Q_1^2-\bar{Q}_1^2)=-2mN
\lb{r-s}
\ee
where $N=0,-n$ according to the allowed solutions (\r{nN00}) and (\r{nN0n}).
Averaging the first two conditions we get the third one as long as we identify
\be
t=(r+s)/2
\ee
According to the discussion in section~\r{sec.review} $r$ and $s$ have
to be even in order the full conformal dimension to be integer. Making the identifications
\be
\ba{ccc}
\displaystyle \frac{r}{2}=1-L \vspace{.2 cm}\\
\displaystyle \frac{s}{2}=1-\bar{L}\vspace{.2 cm}
\ea
\ee
we retrieve the mass shell condition (\r{mshell}). And (\r{r-s}) becomes
the spin condition (\r{spin}).

Below we explore the possible intermediate processes by performing the sums over 
each of the $l_i$'s. 
We obtain a chain of conditions $\phi_i \in \mathbb{Z}$ which allow us to build
the full set of possible diagrams contributing to the transitions. 
Starting with $i=1$ we find
\be
n(N'_1-N_1)+2N_1N'_1=0
\ee
which has two possible solutions
\be
\ba{lrcccr}
a\ \ \ \ \ &N_1&=&0&=&N'_1\vspace{.2 cm}\\
b          &N_1&=&-n&=&-N'_1
\ea
\lb{ab}
\ee
that we name $a$ and $b$. 

For $i=2$ we have
\be
n(N'_1-N_1)+n(N'_2-N_2)+2N_1N'_1+2N_2N'_2+2N_1N'_2+2N_2N'_1=0
\ee
Choosing $a$ for $N_1$ and $N'_1$ we have again the same previous solutions
for $N_2$ and $N'_2$. Choosing $b$ we obtain again $a$ or a new solution
\be
\tilde{b}\ \ \ \ \ N_2=n=-N'_2\\
\lb{bb}
\ee
that we name $\tilde{b}$. If we choose $\tilde{b}$ we will find for $i=3$ an $a$ or $b$
solution. So no new solutions emerge.

Note that a solution $a$ will reduce the
diagram to one of type $i-1$, we can join the linkings before and after that monopole 
since it doesn't change the charge. Without loss of generality we will disregard the
$a$ solution in the following discussion.

By induction
the $i$th condition reads then
\be
n(1-2\#b+2\#\tilde{b})(N'_i-N_i)+2N_iN'_i=0
\ee
with the restriction that $\#b-\#\tilde{b}$ takes only the values $0$ and $1$.
Then we will obtain a chain of alternating solutions $b\tilde{b}\ldots b\tilde{b}b\tilde{b}\ldots$
as the only possible solution. Some diagrams are presented in figure~\r{fig2}.

The above results are consistent with global charge conservation. And moreover they show that
local charge violation is quite restricted. given some charge $q$ in the first boundary
it can change to $\bar{q}$ and return to its previous form $q$ again as many times as
it wants but it can become nothing else.

\fig{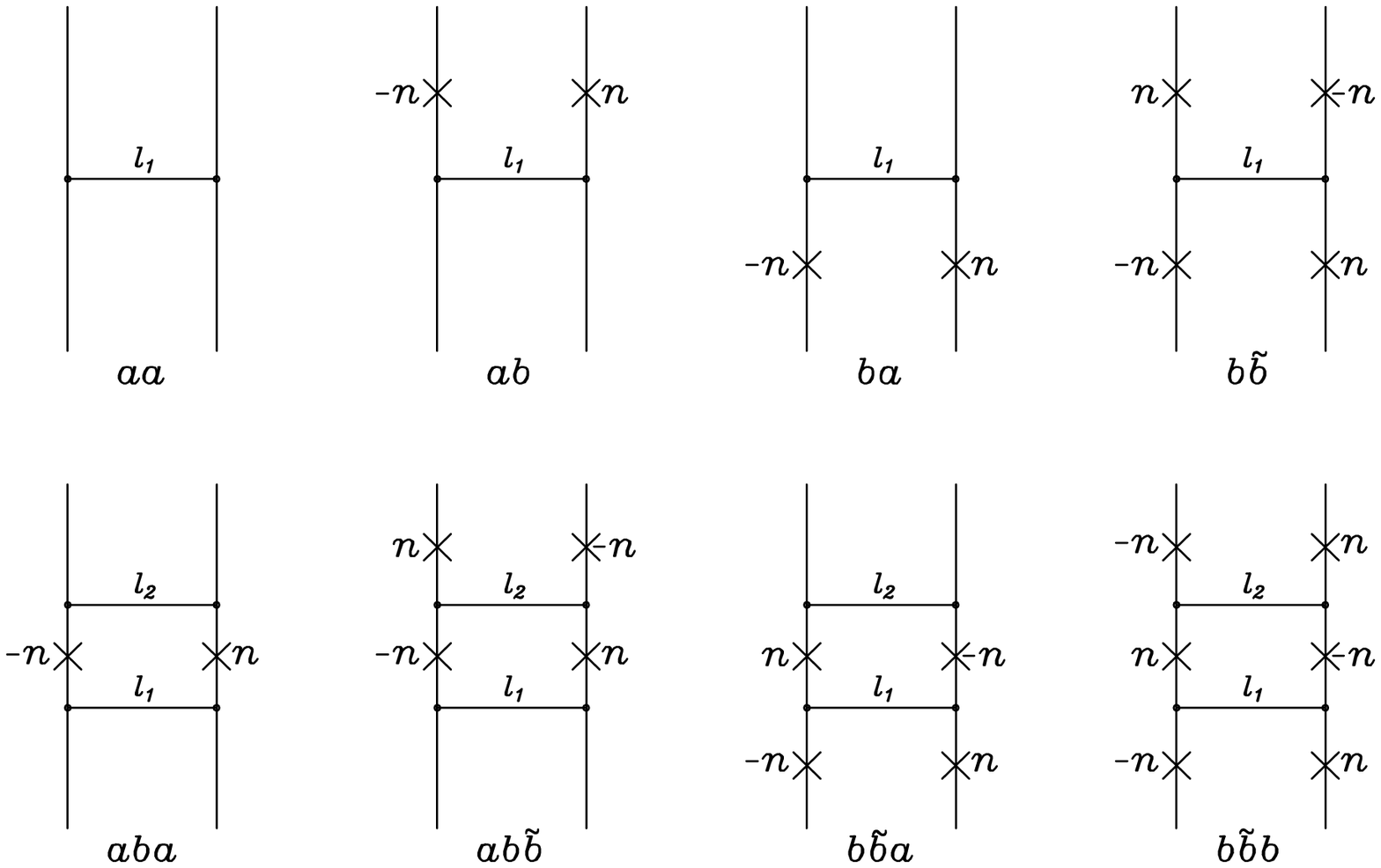}{Some of the possible diagrams for two charge
propagation, $i=2$ and $i=3$.}{fig2}

Lets go on to consider now 3 charges propagating from one boundary to the other
(or equivalently evolving in time)
\be
\ba{ccc}
Q_1&\rightarrow&\bar{Q}_1\vspace{.2 cm}\\
Q_2&\rightarrow&\bar{Q}_2\vspace{.2 cm}\\
Q_3=-Q_1-Q_2&\rightarrow&\bar{Q}_3=-\bar{Q}_1-\bar{Q}_2
\ea
\ee
as pictured in figure~\r{fig1}. The phase is know 
\be
\ba{rcl}
\Phi_3=&\displaystyle
      \left.\frac{2}{k}\,\right\{&Q_1Q_2 l_0 -Q_2(Q_1+Q_2)d_0-Q_1(Q_1+Q_2)k_0+ \vspace{.2 cm}\\
      &+&\displaystyle
      \sum_{i=1}^{j-1}\left[\bar{Q}_{1(i)}^2l_i-\bar{Q}_{1(i)}\bar{Q}_{3(i)} d_i-\bar{Q}_{2(i)}\bar{Q}_{3(i)}k_i\right]\vspace{.2 cm}\\
      &+&\displaystyle\left.\bar{Q}_1\bar{Q}_2 l_j -\bar{Q}_2(\bar{Q}_1+\bar{Q}_2)d_j-\bar{Q}_1(\bar{Q}_1+\bar{Q}_2)k_j\ \right\}
\ea
\ee
where we defined
\be
\ba{rclcrcl}
\bar{Q}_{1(i)}&=&\displaystyle Q_1+\frac{k}{2}\sum_{a=1}^iN_i&\ \ \ \ \ \ &\bar{Q}_{1}&=&\displaystyle Q_1+\frac{k}{2}N \vspace{.2 cm}\\
\bar{Q}_{2(i)}&=&\displaystyle Q_2+\frac{k}{2}\sum_{a=1}^iM_i& &\bar{Q}_{2}&=&\displaystyle Q_2+\frac{k}{2}M \vspace{.2 cm}\\
\bar{Q}_{3(i)}&=&\displaystyle Q_1+Q_2-\frac{k}{2}\sum_{a=1}^iN'_i& & \bar{Q}_{3}&=&\displaystyle Q_1+Q_2-\frac{k}{2}N'
\ea
\ee
Now we can choose to factor out some combination of $l_0$, $d_0$, $k_0$, $l_j$, $d_j$ and $k_j$.
But it is now clear from the previous discussion of two charges propagation
what factorizations we should look for. So we are only going to eliminate all
the $0$'s, all the $j$'s or some combination of $0$'s and $j$'s for the three linkings.
Defining the total linking numbers $l=l_0+\sum l_i+l_j$, $d=d_0+\sum d_i+d_j$
and $k=k_0+\sum k_i+k_j$ we can replace the $l_0$, $d_0$ and $k_0$
dependence and obtain the phases
appearing in the correlation of three fields given in~(\r{OPE3}) for
$z_{12}$, $z_{23}$ and $z_{13}$ on $l$, $d$ and $k$ respectively.
Considering extra non compact gauge fields with charges $p_1$, $p_2$ and $p_3$
and summing over the previous variables we obtain the conditions that the combinations
of conformal dimensions in (\r{OPE3}) must be integer. The conformal dimensions
are to be read considering the extra non compact gauge fields and charges
$\Delta=Q^2/k+p^2/k'$ and $\bar{\Delta}=\bar{Q}^2/k+p^2/k'$. From now on let us use
this definition of conformal dimensions.
As discussed in section~\r{sec.review} from the point of view of the boundary CFT's these
factors must indeed be integer to insure that the three point OPE are single valued.

In terms of the individual vertices it is not so clear what these conditions mean.
Without loss of generality and in order to clarify it let us
replace the sums over $l$, $d$ and $k$ by the sums over
three new variables $l_+$, $d_+$ and $k_+$ such that $l=l_++k_+$, $d=d_++l_+$ and $k=k_++d_+$.
Summing over these new variables one obtains the following conditions
\be
2\Delta_1,\ 2\Delta_2,\ 2\Delta_3\ \in\mathbb{Z}
\lb{l03}
\ee
In a similar way we may change $l_j$, $d_j$ and $k_j$ dependence by
$l_+$, $d_+$ and $k_+$ and upon summation obtain the respective conditions
\be
2\bar{\Delta}_1,\ 2\bar{\Delta}_2,\ 2\bar{\Delta}_3\ \in\mathbb{Z}
\lb{lj3}
\ee
Or using the definition (\r{lpm}) for $l^\pm_0$, $d^\pm_0$ and $k^\pm_0$
and replacing $l^+_0$, $d^+_0$ and $k^+_0$ dependence by $l_+$, $d_+$ and $k_+$
in the same way as before, after performing the sums
in these last variables, we get the conditions
\be
\Delta_1+\bar{\Delta}_1,\ \Delta_2+\bar{\Delta}_2,\ \Delta_3+\bar{\Delta}_3\ \in\mathbb{Z}
\lb{l+3}
\ee
Similarly to the previous discussion for two charge propagation these
conditions turn out to be equivalent to the mass shell and spin conditions of
string theory. 

The fundamental differences arrive in the remaining conditions.
Taking the first two cases~(\r{l03}) and~(\r{lj3}), where we replaced the
$0$'s and $j$'s linking number sums, and summing over
$(l_{+j},d_{+j},k_{+j})$ and $(l_{+0},d_{+0},k_{+0})$
we obtain in both cases three conditions.
The first two are redundant, they correspond to (\r{int0}) which has been obtained from the two
charges propagation, the third one is new
\be
\frac{4}{k}\left(Q_1Q_2-\bar{Q}_1\bar{Q}_2\right) \in \mathbb{Z}
\lb{I3a}
\ee
Nevertheless the solutions for
irrational $k$ end up by being of the same kind as before.
There are two solutions
\be
\ba{ccc}
\left\{\ba{ccc}N&=&0\\M&=&0\ea\right.&\ \ \ &\left\{\ba{ccc}N&=&-n_1\\M&=&-n_2\ea\right.
\ea
\lb{N3}
\ee
For (\r{l+3}) case, where we replaced the sum over $+$'s linking number,
we  obtain, upon summation over $(l_{+0}^-,d_{+0}^-,k_{+0}^-)$, the conditions
two conditions~(\r{even}) which were obtained in the two charges case.
The third one is again new
\be
\frac{2}{k}\left(Q_1Q_2-\bar{Q}_1\bar{Q}_2\right) \in \mathbb{Z}
\lb{I3}
\ee

The solutions for the full diagrams are very similar to the ones of two charges.
With the charges oscillating simultaneous between $Q_1$, $Q_2$
and $\bar{Q}_1=Q_1-n_1k/2$, $\bar{Q}_2=Q_2-n_2k/2$. For four charges there are no new conditions.

After all these algebra let us summarize what we obtained. We started with a theory
which, at the perturbative level, has the allowed charges of the form $Q = m + (k/4)n$. 
We showed, under certain assumptions that, if the non-perturbative processes are taken into
account, the charge spectrum is modified (restricted) quite drastically. The allowed
left/right pair of charges are in one-to-one correspondence with the processes which
have non-vanishing quantum amplitudes. All these processes can be organized in a lattice
$\Gamma$ with elements $l=(Q,\bar{Q}=Q+(k/2)N)$ with $N$ either $0$ or $-n$.
Moreover a Lorentzian product $\circ$ of signature ($+$,$-$) emerge naturally
from~(\r{int0}),~(\r{even}) and~(\r{I3}) defined as
\be
l\circ l'=\frac{2}{k}(QQ'-\bar{Q}\bar{Q}')
\lb{lorprod}
\ee
And we can go even further and extract very important properties of the lattice,
it is {\bf even} due to~(\r{even})
\be
l\circ l=\frac{2}{k}(Q^2-\bar{Q}^2)\in2\mathbb{Z}
\lb{EVEN}
\ee
and {\bf integer} due to~(\r{I3}).
\be
l\circ l'=\frac{2}{k}(QQ'-\bar{Q}\bar{Q}')\in\mathbb{Z}
\lb{INT}
\ee
Note that~(\r{I3a}) is redundant since it is necessarily obeyed if~(\r{I3}) is.

By inspection, for $N=-n$ the lattice is {\bf self-dual} as well. But not for $N=0$.
In the next section we explain how to exclude $N=0$ case and obtain what we seek.
The lattice is then exactly the one for the
bosonic string theory with one compact dimension. It is the Narain lattice
for compactification in $S^1$.
For toroidal compactification of several dimensions is enough to consider several
compact $U(1)$. This situation is well described in section~\r{sec.lat}.

\section{Boundary conditions \lb{sec.bc}}

We still need to understand how we can select between $N=0$ and $N=-n$ cases. The key is
to consider different combinations of the boundary conditions introduced in~(\r{bc}).
Further, to obtain a CFT in the boundary (whether it is chiral or not) as an effective 2D theory
of the 3D TMGT we have to identify the two boundaries $\Sigma_L$ and $\Sigma_R$ in some way.
These are the mechanisms that allow us to build several string theories out of the
same $3D$ theory.
Note that this constructions are only possible due to the Maxwell term in the action
and the existence of monopoles.

We learned in the last section that the bulk theory only allows the charge $q= m+kn/4$ to become
$\bar{q}=m-kn/4$ or keep it to be the same charge $q$.
What is actually $n$? Let us go back to (\r{q}), $2 \pi n$ is the flux of
the magnetic field. So upon some kind of boundary identification our theory only admits 
the ones that hold one of the conditions
\be
\ba{lcl}
N=0& &\displaystyle\int_{\Sigma_L}d^2z_LB=\int_{\Sigma_R}d^2z_RB\vspace{.2 cm}\\
N=-n&\ \ \ &\displaystyle\int_{\Sigma_L}d^2z_LB=-\int_{\Sigma_R}d^2z_RB
\ea
\lb{id0}
\ee
Let us consider a map (coordinate transformation) that transforms a vector in $\Sigma_R$ into
another vector in $\Sigma_L$, giving us the identification \textit{rule}.
There will be two kinds of maps. One which maintain the relative orientation
of both $2d$ boundaries, let us call it parallel (${\small//}$), and the other kind
that reverses the relative orientation, let us call it perpendicular ($\perp$).
The names are chosen by the relative identification of the axes
from boundary to boundary as pictured in figure~\r{fig_id01}.

Note that the induced $\tilde{\epsilon}$ does not 
change from boundary to boundary. From the $3d$ point of view we are
reversing time in one boundary and swaping the complex space
coordinates. From the point of view of the bulk nothing special is
happening.

\fig{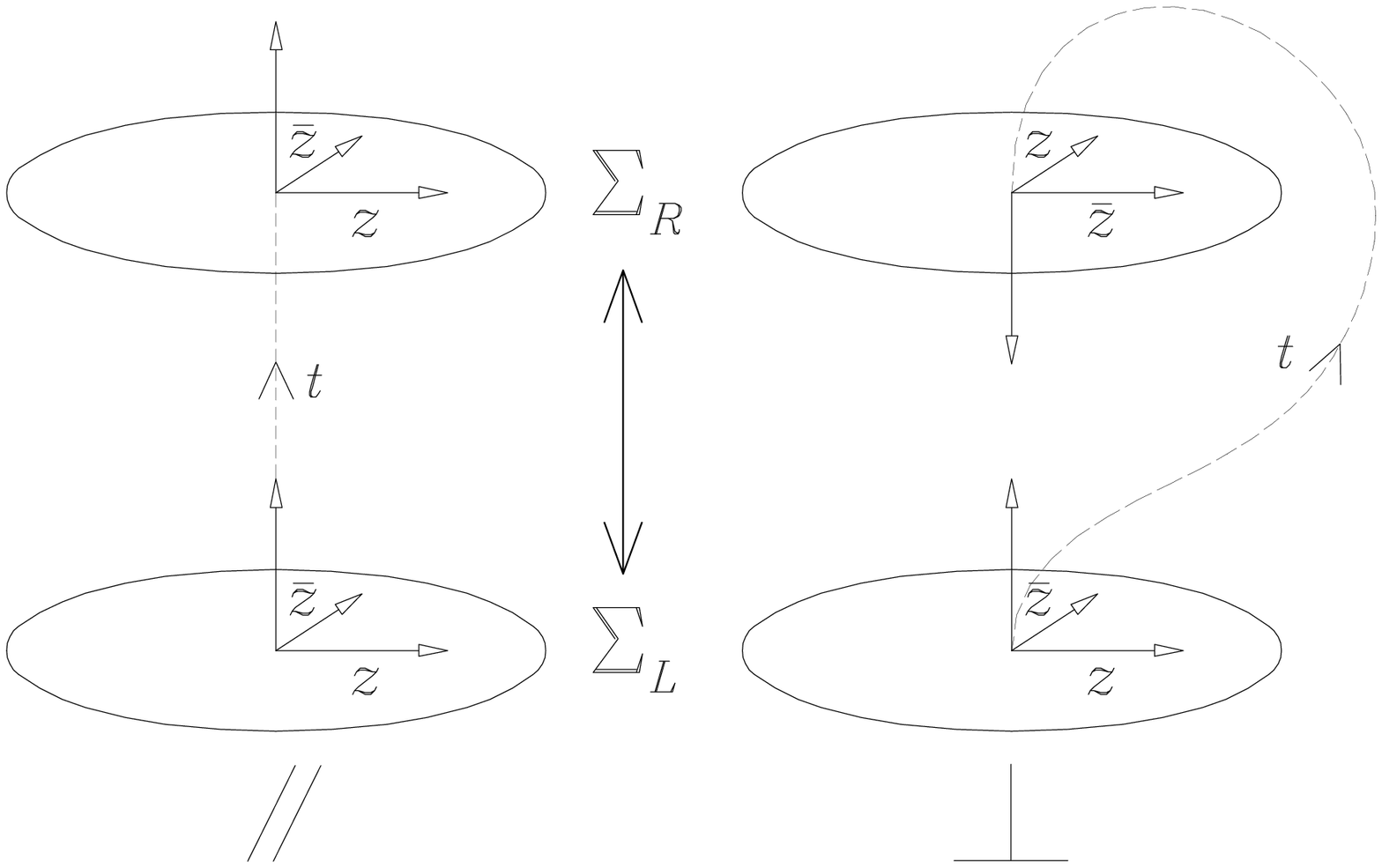}{Boundary identification with the same orientation (parallel; ${\small//}$)
and reversed orientation (perpendicular; $\perp$)}{fig_id01}

Let us define the difference of fluxes as
\be
\delta\phi=\int_R B-\int_L B=
\int_{\Sigma_R}d^2z\epsilon^{ij}\partial_iA_j-\int_{\Sigma_L}d^2z\epsilon^{ij}\partial_iA_j
\ee
Given our boundary conditions we can evaluate this difference explicitly. Let us write
what the magnetic field is for each of the allowed boundary conditions
\be
\ba{lrl}
C.&\phi_C=&-\int\partial_\bz A_z\vspace{.2 cm}\\
\bar{C}.&\phi_{\bar{C}}=&\int\partial_z A_\bz
\ea
\ee
In these equations we took the magnetic field definition
$B=\partial_zA_\bz-\partial_\bz A_z$ and imposed the respective boundary
conditions.
Taking into account the boundary identifications and labeling the
different combinations of boundary conditions accordingly
we compute the flux difference
\be
\ba{lc}
CC_{\small//}.&\delta\phi=0\vspace{.2 cm}\\
CC_\perp.&\delta\phi=-\int(\partial_\bz A_z+\partial_z A_\bz)=-2\pi(2n)\vspace{.2 cm}\\
C\bar{C}_{\small//}.&\delta\phi=-\int(\partial_\bz A_z+\partial_z A_\bz)=-2\pi(2n)\vspace{.2 cm}\\
C\bar{C}_\perp.&\delta\phi=0
\ea
\ee
Note that for parallel type of identifications the fields and
integrals are summed without any relative change. For the
perpendicular type we have to change the space indices of the fields
to their conjugates and the measure in the integral changes 
sign (in one boundary, say the right one).
The results are summarized in figure~\r{fig_id02}.

\fig{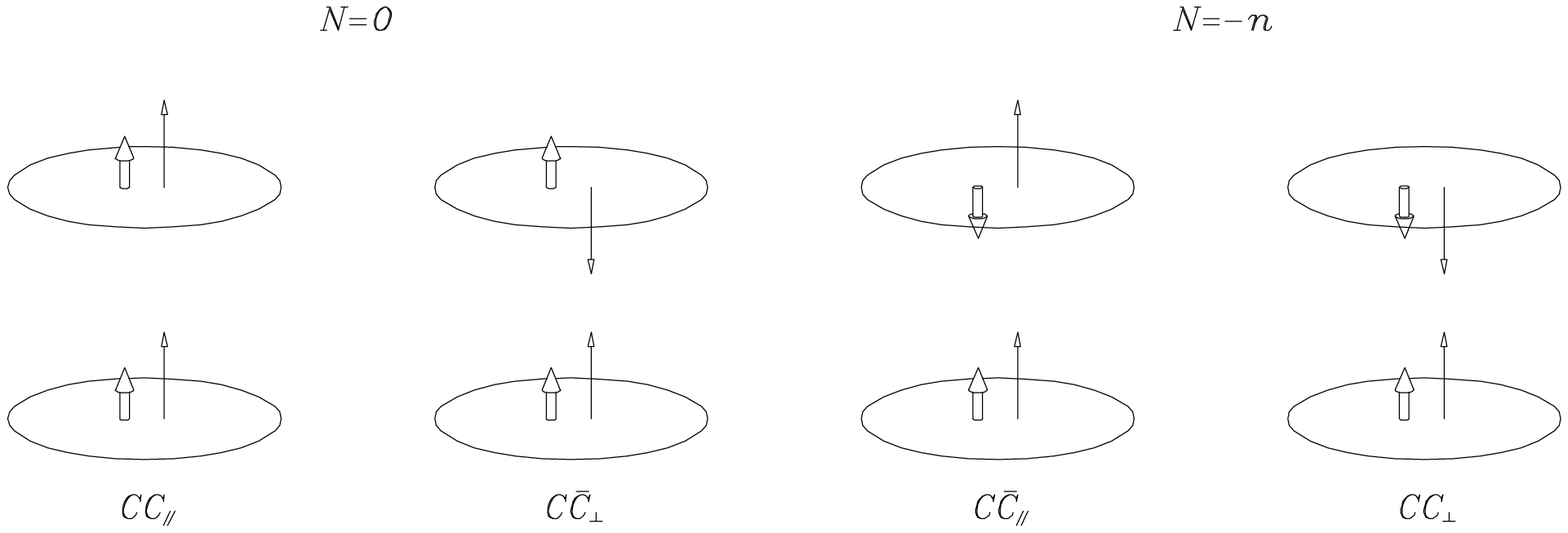}{Boundary identifications for several combinations of boundary conditions.
The fat arrow represents magnetic flux and the thin one boundary orientation.}{fig_id02}

So the conclusion is that
$Q=\bar{Q}$ corresponds to $CC_{\small//}$, $\bar{C}\bar{C}_{\small//}$, $C\bar{C}_\perp$
or $\bar{C}C_\perp$ and $Q-\bar{Q}=(k/2) n$ to
$CC_\perp$, $\bar{C}\bar{C}_\perp$, $C\bar{C}_{\small//}$
or $\bar{C}C_{\small//}$ boundary conditions.
Note that in principle there may exist other choices of the maps which would give
different boundary theories. We leave this to a future work.

Note that the spectrum for $NC$ or $N\bar{C}$ boundary conditions must be obtained
by truncating the spectrum of $CC_\perp$ or $C\bar{C}_{\small//}$. We simply pick $\bar{Q}=0$
for the $N$ boundary, for instance
\be
m=\frac{k}{4}n
\lb{QN}
\ee
and the charges in the other boundary become
\be
Q=\frac{k}{2}n
\lb{QQN}
\ee
Of course for irrational $k$ the only solution for the condition (\r{QN}) is $m=n=0$.
We end up with a very poor and empty theory in the boundary.
In section~\r{sec.lat} it will be clarified in which cases condition (\r{QN}) 
allows some dynamics in the boundary.
Trying to truncate the spectrum of $CC_{\small//}$ or $C\bar{C}_\perp$ we
will set straight away $Q=\bar{Q}=0$ since the charges are equal in
both boundaries killing the hope of finding any dynamics in the
boundaries.

We choose from now on to work with $CC_\perp$ or $C\bar{C}_{\small//}$ type of boundary
conditions since they are the ones which give us the desired spectrum
in the boundaries CFT's. Further, as we just
explained, $N$ boundary conditions can easily be obtained from them.

\section{RCFT's and Fusion Rules\lb{sec.cb}}

A CFT is rational when its infinite set of primary fields (vertex operators) can be organized
in a finite number $N$ of families usually called \textit{primary blocks}. In each of those
blocks we can choose one minimal field that is the generator of that family. There is
an algebra between these families, or if we want, between the $N$ minimal fields called
the fusion algebra. The fusion rules define this second algebra of fields.

In what follows we are only going to discuss the holomorphic part of
the c=1 RCFT of a bosonic field $\phi$ living in a circle of radius $R=\sqrt{2p'/q}$.
Here $p'$ and $q$ are integers. The vertex operators are
\be
V_{Q_c}=\exp(2\pi i Q_c \phi)\ \ \ \ \ \ Q_c=\frac{r}{R}+s\frac{R}{2}\ \ \ \ \ \ r,s\in \mathbb{Z}  
\ee
Where $Q_c$ are the charges (or momenta) of the theory. The conformal dimensions of
these vertex operators are $\Delta=Q_c^2/2$. There are $N=2p'q$ primary blocks (or families).
The generators $V_{Q_\lambda}$ of such families are chosen such that their conformal dimensions are the
lowest allowed by the theory. In terms of their charges these are
\be
Q_\lambda=\frac{\lambda}{\sqrt{N}}\ \ \ \ \lambda=0,1,2,\ldots,N-2,N-1
\ee
We are going to call them from now on primary charges.
In this way $Q_\lambda$ runs from $0$ to $\sqrt{N}-1/\sqrt{N}$.
The remaining fields of the theory are obtained by successive products
of the generators. The charges in a family $\lambda$ are
\be
Q_{\lambda,L}=Q_\lambda+L\sqrt{N} \ \ \ \ \ \ L\in \mathbb{Z}
\ee
The generators form the fusion algebra given by the fusion rules
\be
Q_{\lambda}+Q_{\lambda'}=Q_{(\lambda+\lambda')\mod N}
\ee
We took the liberty to express it in terms of the charges. Formally it is expressed in terms
of fields or vertex operators $V_{\lambda}\times V_{\lambda'}=V_{(\lambda+\lambda')\mod N}$. 
This simply means we pick two primary charges
out of two families and add them. We will obtain a charge from a different family, but
not necessarily the primary one.
The fusion rules pick the primary charge of that new family.
For further details on these subjects see~\cite{FMS}.

For our normalization we have the following relations 
\be
k=2R^2\ \ \ \ Q=R Q_c
\ee

In what follows we are going to explain how all this emerges with some naturalness
from the bulk theory.
Take then $k=2p/q$ to be a rational number. Note that in the previous discussion $2p'=p$,
so take even $p$.
By inspection we check that the $(m,n)$ space is divided in diagonal bricks
containing $p\times q$ charges. They are distributed in diagonal layers of
identically valued bricks as symbolically pictured in figure~\r{figbricks}.
To check this \textit{diagonal structure} explicitly take a generic charge labeled by the
pair $(m,n)$, it can be represented by any other pair $(m-pn'/2 ,n+q n')$
\be
Q=m+\frac{p}{2q}n=m-\frac{p}{2}\ n'+\frac{p}{2q}(n+q\ n')
\lb{Qp}
\ee
This simply represents a diagonal translation of $n'$ bricks
in the figure.

\fig{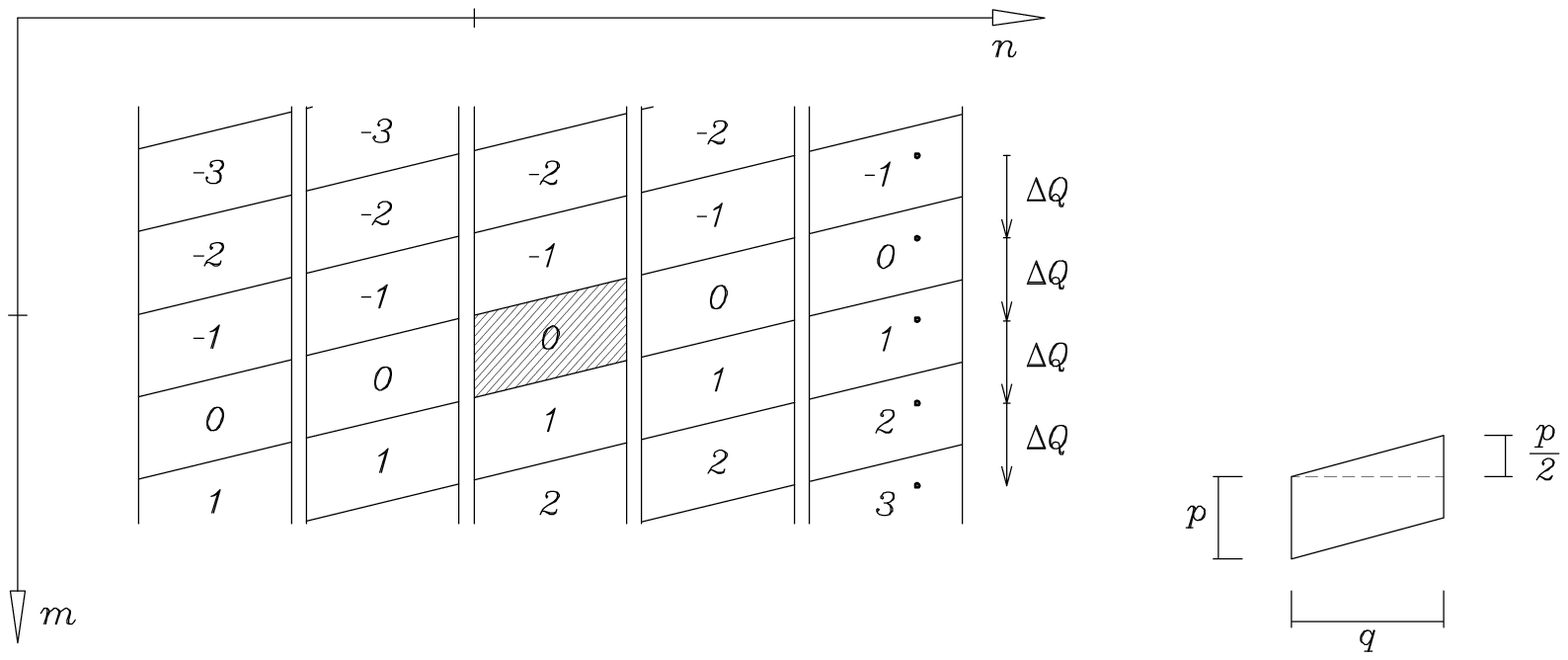}{$(m,n)$ structure for even $p$. The dashed region
is the primary brick $(r,s)$, some examples are presented in figure~\r{figbrick}.
The numbers inside the blocks represent $L \Delta Q$ shift of the charge
values as given by (\r{DQ}). Blocks with the same $L$ have the same charge entries. In each
brick there is exactly one element of level $L$ belonging to each
family.}{figbricks}

Note that this choice of the brick shape is not unique, we could take some other choice
as long as it has dimension $pq$ (e.g. parallelogram of sides $p$ and $q$)
and the following results would hold nevertheless. The reason for this 
particular choice
is in order to get a direct parallel with the usual result of RCFT's.
Further we consider one of such bricks to be
the primary brick. It corresponds to the minimal charges (chosen to be positive) allowed
by the theory built out of the lowest pairs of integers $(r,s)$
\be
Q_{rs}=r+\frac{p}{2q}s=\frac{\lambda}{q}\ \ \ \ \ \ \lambda=rq+\frac{p}{2}s=0,1,2,\ldots,pq-2,pq-1
\ee
Examples for some values of $k$ are given in figure~\r{figbrick}. The geometric rule
to organize the charges inside the brick is to order them in ascending order
by their distance to the upper diagonal line of slope $k/4$.

\fig{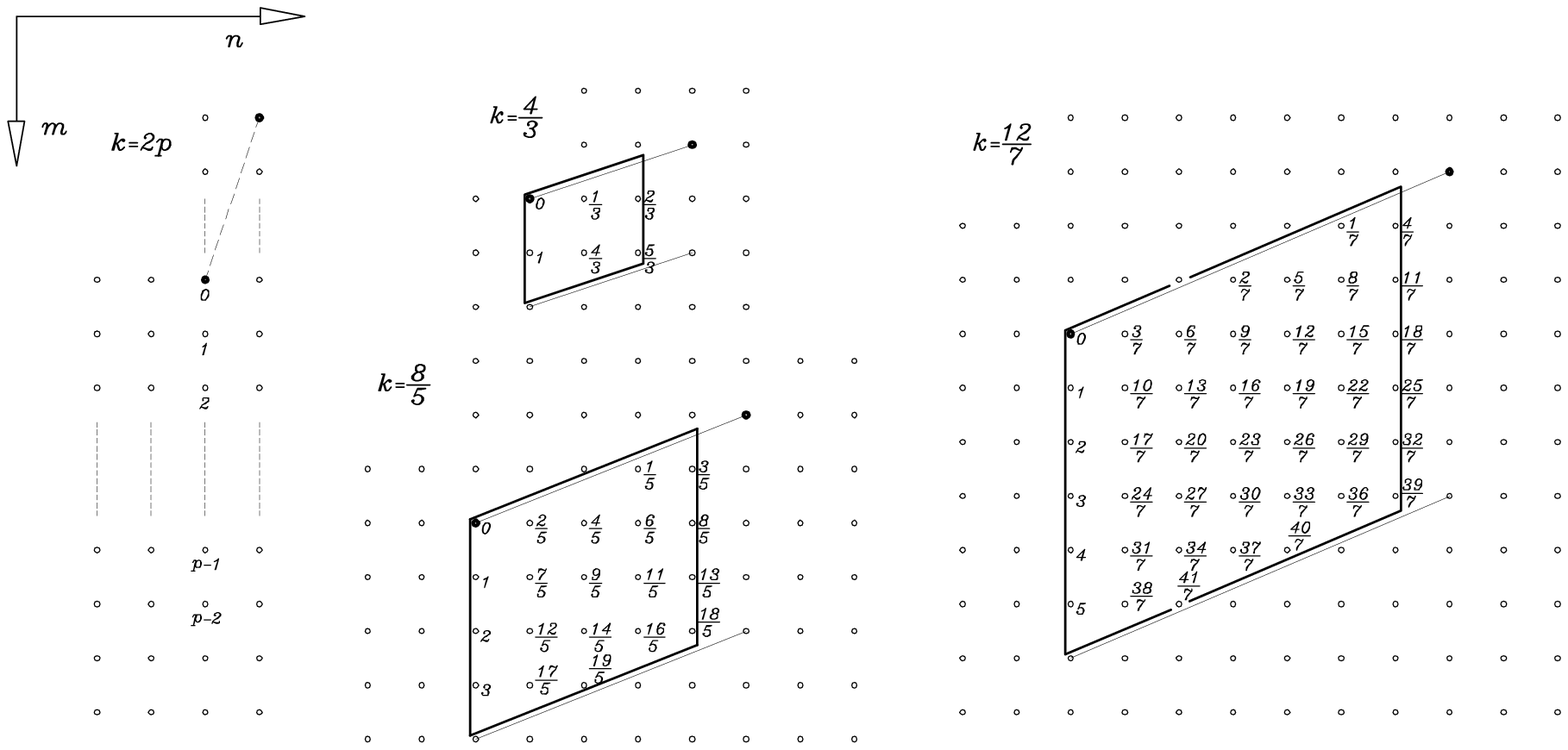}{Primary charges distribution in the $(m,n)$ plane, they
constitute the primary brick $(r,s)$. The rule to order the charges is by  their distance
to the (thin) line connecting the charges $Q=0$ (filled dots) of slope $k/4=p/2q$.}{figbrick}

So far we just reproduced the well know charge structure of RCFT's, it is
now time to return to the bulk theory and justify it.
Take again condition (\r{even}), replacing $k$,
it reads now
\be
\frac{p}{2q}\left(nN-N^2\right) \in \mathbb{Z}
\ee
which can be reexpressed as
\be
N(n-N)=0\mod q
\lb{evenrac}
\ee
The solutions for (\r{evenrac}) can be easily computed to be
\be
\ba{ccc}
N&=& 0\mod q\vspace{.2 cm}\\
N&=&-n+0\mod q
\ea
\ee
which are equivalent to (\r{int0}) and (\r{int01}).
There is one important lesson to take from this result,
besides the previous allowed monopole-instanton
process $n\rightarrow-n$ which is charge dependent, there is a new
charge independent one:
\be
\Delta Q=\pm p
\lb{DQ}
\ee
This is actually the physical process that spans each of the families!
To obtain the charges of some family in terms of $m$ and $n$ we can think on shift either
$m\rightarrow m+pL$ or $n\rightarrow n+2qL$. This is due to an infinite degeneracy
in the $(m,n)$ plane of the charges values expressed by (\r{Qp}).
Figure~\r{figcharges} presents the distribution of charges for $k=2p$. In this
simplified case the families are simply organized along the $m$ axis for even
values of $2n$. This structure is repeated by shifts on the $m$ axis of value $p$.

\fig{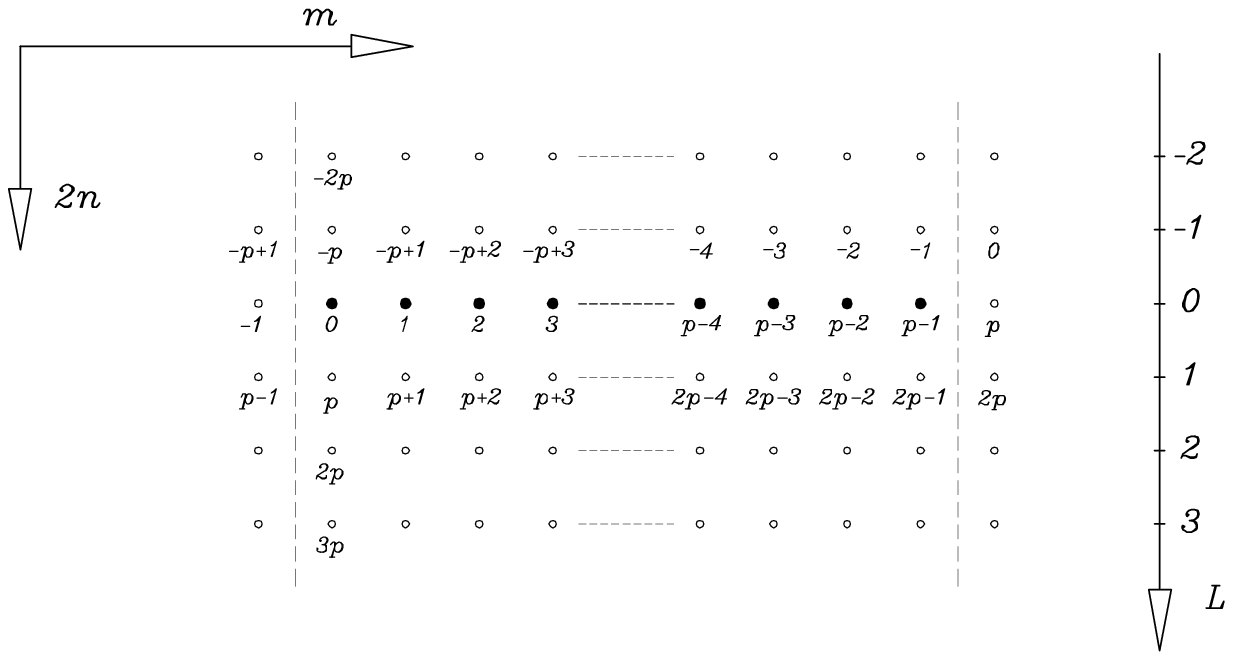}{Charge structure for $k=2p$. Only even values are considered
in the $n$ axis}{figcharges}

For odd $p$ all the structure is similar but with bricks of dimension $p\times 2q$, 
take again a generic charge Q
\be
Q=\frac{p\ n}{2q}+m=\frac{p\ (n+2\ n')}{2q}+m_p-p\ n'
\lb{Qpp}
\ee
Because $p$ is odd, same valued charges correspond to pairs related by $(m,n)\rightarrow(m-pn',n+2qn')$.
The slope of the brick is the same $k/4$ but they are twice bigger in the $n$ direction.
All the rest works in the same way. This would correspond to a boson
field living on a circle of radius $R=\sqrt{p/q}$.

\section{Lattices and Heterotic String \lb{sec.lat}}

Considering an action of the type (\r{S}) with compactified $U(1)^D$
gauge group (replacing $k$ by a generic tensor $K_{IJ}$) we
have
\be
S=\int_{\cal{M}} d^2z\,dt\left[-\frac{\sqrt{-g_{(3)}}}{4\gamma}F_I^{\mu\nu}F^I_{\mu\nu}+
\frac{G_{IJ}+B_{IJ}}{8\pi}\epsilon^{\mu\nu\lambda}A^I_\mu\partial_\nu A^J_\lambda\right]
\lb{SK}
\ee
$G_{IJ}$ and $B_{IJ}$ stand for the symmetric and antisymmetric parts
of $K_{IJ}$.
Using the same procedure outline in section~\r{sec.review} we
build the monopole-instanton which induces a charge change of
\be
\Delta Q_I=\frac{G_{IJ}}{2}N^J
\lb{DDQQ}
\ee
Using the Schr\"{o}dinger picture we get the charge spectrum
\be
Q_I=m_I+\frac{K_{IJ}}{4}n^J
\ee
Note that the antisymmetric part $B_{IJ}$ is present in the charge but
not in the monopole effects. This is due to the fact that it is
manifested only at the level of the boundary. Note that in the action
the term corresponding to $B_{IJ}$ is a total derivative and can be
completely integrated out to the boundary
\be
\int_{M}B_{IJ}\epsilon^{\mu\nu\lambda}A^I_\mu\partial_\nu A^J_\lambda=\frac{1}{2}\left[-\int_{\Sigma_L}B_{IJ}\tilde{\epsilon}^{ij}A_i^IA_j^J+\int_{\Sigma_R}B_{IJ}\tilde{\epsilon}^{ij}A_i^IA_j^J\right]
\ee
Let us analyze first the case for $B_{IJ}=0$. The previous discussion
of section~\r{sec.bc} follows in the same fashion. Choosing $CC_\perp$
or $CC_{\small//}$ boundary
conditions we have the desired relative spectrum in each boundary,
that is, $\vb{Q}=\vb{m}+K\vb{n}/4$ and $\bar{\vb{Q}}=\vb{m}-K\vb{n}/4$. This means that every
monopole contribution of the form (\r{DDQQ}) has exactly $N^J=-n^J$.

What about if $B_{IJ}\neq 0$?
Let us return to our boundary identifications,
For $CC_\perp$ the measure of the integrals in opposite
boundaries change their relative sign, say $\int_Rd^2z\rightarrow-\int_Rd^2z$,
due to the $2d$ measure changing sign. The fields in one boundary (say $\Sigma_R$)
are swaped, $A_i\leftrightarrow A_j$. Note that we are not changing
neither $B_{IJ}$ nor $\tilde{\epsilon}^{ij}$ in the right integral, they
are induced from the bulk and do not change by the boundary identifications.
Thus this transformation has no effect in the $B_{IJ}$ term
\be
\int_{\Sigma_L}B_{IJ}\epsilon^{ij}A_i^IA_j^J\rightarrow \int_{\Sigma_L}B_{IJ}\epsilon^{ij}A_i^IA_j^J
\ee
For $CC_{\small//}$ nothing changes either.
This means that $B_{IJ}$ term does not change sign under any of our
boundary identifications.

So we obtain the left/right spectrum
\be
\ba{c}
\displaystyle Q_I=\frac{G_{IJ}+B_{IJ}}{4}n^J+m_I\vspace{.2cm}\\
\displaystyle \bar{Q}_I=\frac{-G_{IJ}+B_{IJ}}{4}n^J+m_I
\ea
\ee
such that the charge difference $Q_I-\bar{Q}_I=G_{IJ}n^J/2$
is indeed the monopole contribution.

We obtain a lattice $l=(\vb{Q},\vb{\bar{Q}})$
with the Lorentzian product of signature $(\vb{+},\vb{-})$ defined as
\be
l\circ l'=2G^{(-1)IJ}(Q_I{Q'}_J-\bar{Q}_I{\bar{Q}'}_J)
\lb{lorprod2}
\ee
where $G^{-1}$ stands for the inverse of $G_{IJ}$.
The signature of the product has $D$ plus and $D$ minus.
The properties of this lattice are the same as the ones for the
previous $D=1$ case and follow in a similar way from the bulk theory
as presented in section~\r{sec.bulk}. The lattice is integer, even and 
self-dual.

Let us analyze which lattices do exist for $CN$ boundary conditions.
As explained before (section~\r{sec.bc}) they can be obtained by
truncating the lattices where $N^J=-n^J$.
Imposing then $CN$ is equivalent to truncating the lattice
by choosing $\vb{\bar{Q}}=0$. Similarly to (\r{QN}), this means we are selecting
elements of our lattice such that
\be
m_I=\frac{G_{IJ}-B_{IJ}}{4}n^J
\lb{QQN2}
\ee
This sublattice has elements $l=(\vb{Q},0)$ with $\vb{Q}$ built out of $\vb{n}$ and $\vb{m}$ obeying (\r{QQN2}).
Once $\bar{\vb{Q}}=0$ the Lorentzian product (\r{lorprod2})
becomes, for this particular sublattice, simply $2G^{IJ} Q_I Q_J$. So it becomes Euclidean
for this particular choice of elements. But the properties of being
even, integer and self dual are inherited from the full lattice.
It is a known fact that the only
even integer self dual Euclidean lattices are of dimension
$0\mod 8$. Of dimension $8$ there is only one, $\Gamma^8$, the root lattice of $E_8$.
From (\r{QQN2}) we can derive the spectrum of allowed $\vb{Q}$'s
\be
Q_I=\frac{G_{IJ}}{2}n^J
\lb{QN2}
\ee
Note that the resulting spectrum is independent of $B_{IJ}$.
Due to the spectrum (\r{QN2}) and by closure of the lattice under the
defined product (\r{lorprod2}) we are obliged not to restrict $\vb{n}$.
Anyhow, as stated before, the charge lattice must be the root lattice
of $E_8$, it is clear that
the only choice that holds this result is $G=2C$. $C$ is the standard
Cartan matrix of this group (the diagonal elements have the value $2$
and the off diagonal $-1$). The corresponding Dynkin diagram is pictured in figure~\r{fige8}.
Let us return to (\r{QQN2}). It must be a realizable condition for every $\vb{n}$,
Thus we have no choice but to impose $G_{IJ}-B_{IJ}=0\mod 4$.

We are left with
\be
\ba{rclc}
G_{IJ}&=&2C_{IJ}& \vspace{.2 cm}\\
B_{IJ}&=&(2+4 r)C_{IJ}&J>I,r\in\mathbb{Z}
\ea
\ee

\fig{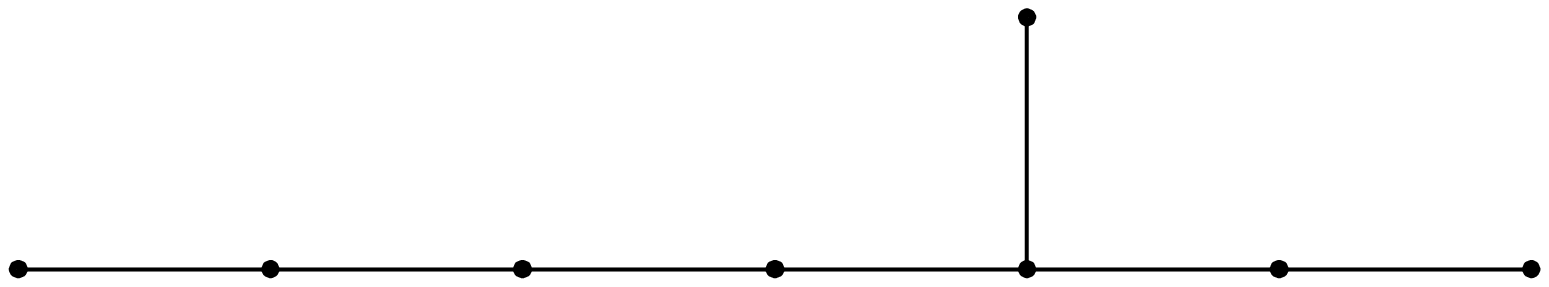}{Dynkin Diagram of $E_8$}{fige8}

There are two lattices of dimension $16$ obeying these properties:
$\Gamma^8\times\Gamma^8$ with elements in the root lattice of
$E_8\times E_8$ and $\Gamma^{16}$. We retrieve the well known results of 
string theory. Note that in the first works \cite{GHMR_1} of non interacting Heterotic string there
is no mention to the tensor $B_{IJ}$ (the free spectrum depends solely 
on $G_{IJ}$). Only after the introduction of
interactions and an effective action for the theory \cite{GHMR_2} it becomes clear that 
there is a need for $B_{IJ}$.
For $r=0$ we retrieve the result stated in \cite{R_1} and references
there in up to normalization of $1/4$ on $K_{IJ}$.

\section{Conclusions \lb{sec.conc}}

Most of the constraints on string theories exist because of their intimate relations to $2D$ geometry,
$2D$ conformal field theories and their symmetries. One of these constraints is the modular 
invariance which puts severe restrictions on the spectrum of the theories. 
Modular invariance, as employed by Narain in the context of toroidal compactification, reduces the
physical problem of finding the allowed momenta and winding modes in string theory
to the problem of constructing lattices on a pseudo-Euclidean space ${\mathbb{R}}^{d_R,d_L}$ on 
which physical invariants are modular
invariant functions. These lattices turn out to be
\textit{self-dual}, \textit{integer} and
\textit{even} lattices.

In this paper we have reproduced the results of Narain compactification  
starting from $3D$ gauge dynamics without referring to modular invariance at all. 
The fact that $3D$ topologically
massive gauge theory carries full information about the lattices
of string theory is quite fascinating. Our approach and the modular invariance construction are compatible
and moreover as we will comment on it below they are logically connected. But there is still an element of 
surprise here because of the crucial difference between the ideology of these two approaches. 
To see this let us summarize what we have obtained. 
We started with a TMGT defined on a three manifold with two disconnected boundaries. Gauge degrees of freedom
became dynamical on the boundaries generating chiral CFT's. Each boundary of the three manifold
is interpreted as a ``chiral worldsheet'' of string theory, meaning that the left and
right modes are physically separated. This is the frame work of Topological Membrane theory. 
We have considered the compact Abelian theory which has a discrete charge
spectrum of the form $\vb{Q} =\vb{m} + k \vb{n}/4$.  A particle with charge $\vb{Q}$
is inserted at one boundary and it travels through the bulk interacts with all possible charge violating instantons
on its path and links with other charges and emerges as a new charge  $\bar{\vb{Q}} = \vb{m} - k \vb{n}/4$ on the other boundary. 
The path of the charged
particle is represented by a Wilson line in the bulk theory. After taking care of  all the linkings
and instanton interactions it is quite an amusing result to obtain that the emerging
charge in the other boundary is of the $\bar{Q}$ form and nothing else. 
The Aharonov-Bohm phases of the linkings interfere in a way that  for a particle which does not have the charge $\pm \bar{Q}$
there is zero probability to emerge in the other boundary.
Moreover a natural self-dual Lorentzian lattice
structure emerges from the linkings of the Wilson lines as it was shown in Section~\r{sec.bc}. 
The connection between our approach and the modular invariance arguments was not worked out in this paper but  
naively one can see that modular transformations in the boundaries will yield linkings of the Wilson lines in the bulk.
This is why we have gotten the same results.

There are a couple of important things which need to be addressed in future.
One of them is the bulk interpretation of A-D-E classification~\cite{Z_1} of
modular invariant partition functions in the boundary. An other important issue is to study
non-orientable surfaces which appear for non-oriented strings. We hope to address
these issues in forthcoming publications.\\

\noindent {\bf Acknowledgments}\\We would like to thank Arshad Momen and Martin
Schvellinger for useful discussions. PCF would also like to thank Alex Kovner for
many useful discussions and Gerald Dunne for some remarks on the manuscript.
The work of PCF is supported by PRAXIS XXI/BD/11461/97 grant from FCT (Portugal).
The work of IK and BT is supported by PPARC Grant PPA/G/0/1998/00567.
The work of IK is also supported by EU Grant FMRXCT960090.


\begin{thebibliography}{99}
\bibitem{IK_0}   I. I. Kogan, Phys. Lett. {\bf B231} (1989) 377
\bibitem{KC_0}   S. Carlip and I. I. Kogan, Mod. Phys. Lett. {\bf A6}, No 3 (1991) 171-181
\bibitem{W_1}    E. Witten, Commun. Math. Phys. {\bf 121} (1989) 351-399
\bibitem{MS_1}   G. Moore and N. Seiberg, Phys. Lett. {\bf B220} (1989) 422
\bibitem{MS_2}   S. Elitzur, G. Moore, A. Schwimmer and N. Seiberg, Nucl. Phys. {\bf B326} (1989) 108\\
                 M. Bos and V. P. Nair, Phys. Lett. {\bf B223} (1989) 61; Int. J. Mod. Phys. {\bf A5} (1990) 959\\
                 J. M. F. Labastida and A. V. Ramallo, Phys. Lett. {\bf B227} (1989) 92; {\bf B228} (1989) 214\\
                 W. Ogura, Phys. Lett {\bf B229} (1989) 61
\bibitem{holog}  G 't Hooft, \textit{Salamfest} (1993) 284-296 (e-Print Archive: gr-qc/9310026)\\
                 L. Susskind, J. Math. Phys. {\bf 36} (1995) 6377-6396 
\bibitem{KS_1}   S. Carlip and I. I. Kogan, Phys. Rev. Lett. {\bf 64} (1990) 148; Phys Rev. Lett. {\bf 67} (1991) 3647\\
                 I. I. Kogan, Phys. Lett. {\bf B256} (1991) 369; Nucl. Phys. {\bf B375} (1992) 362\\
                 G. Amelino-Camelia, I. I. Kogan and R. J. Szabo, Nucl. Phys. {\bf B480} (1996) 413-456; Int.J.Mod.Phys. {\bf A12} (1997) 1043-1052\\
                 I. I. Kogan and R. J. Szabo, Nucl.Phys. {\bf B502} (1997) 383-418\\
                 I. I. Kogan, A. Momen and R. J. Szabo, JHEP 9812 (1998) 013
\bibitem{N_1}    K. S. Narain, Phys. Lett. {\bf B231} (1989) 377\\
                 K. S. Narain, M. H. Samardi and E. Witten, Nucl. Phys. {\bf B279} (1987) 369
\bibitem{GHMR_1} D. Gross, J. Harvey, E. Martinec and R. Rohm, Nucl. Phys. {\bf B256} (1985) 253
\bibitem{GHMR_2} D. Gross, J. Harvey, E. Martinec and R. Rohm, Nucl. Phys. {\bf B267} (1986) 75
\bibitem{D_1}    R. Jackiw and S. Templeton, Phys. Rev. {\bf D23} (1981) 2291\\
                 J. Schonfeld, Nucl. Phys. {\bf B185} (1981) 157\\
                 S. Deser, R. Jackiw and S. Templeton, Phys. Rev. Lett. {\bf 48} (1982) 975, Ann. Phys. NY {\bf 140} (1982) 372
\bibitem{Lee}    K. Lee, Nucl. Phys. {\bf B373} (1992) 735
\bibitem{IK_1}   L. Cooper and I. I. Kogan, Phys. Lett. {\bf B383} (1996) 271-280
\bibitem{LC_1}   L. Cooper, D.Phil Thesis, 1997
\bibitem{TEKIN}  P. Ning-Tan, B. Tekin and Y. Hosotani, Phys. Lett. {\bf B388} (1996) 611-620; Nucl. Phys. {\bf B502} (1997) 483-515
\bibitem{IK_2}   L. Cooper, I. I. Kogan and K.M. Lee,Phys. Lett. {\bf B394} (1997) 67-74
\bibitem{KS_2}   L. Cooper, I. I. Kogan and R. J. Szabo, Annals. Phys. {\bf 268} (1998) 61-104
\bibitem{O_1}    W. Ogura, Phys. Lett {\bf B229} (1989) 61
\bibitem{C_1}    S. Carlip, Nucl. Phys. {\bf B362} (1991) 111\\
                 M. Asorey, F. Falceto and S. Carlip, Phys. Lett. {\bf B312} (1993) 477-485\\
                 M. C. Ashworth, Mod. Phys. Lett. {\bf A10} (1995) 2749-2755
\bibitem{AK_1}   E. C. Marino, Phys. Rev. {\bf D38} (1988) 3194\\
                 A. Kovner, B. Rosenstein and D. Eliezer, Nucl. Phys. {\bf B350} (1991) 325\\
                 A. Kovner and B. Rosenstein, Int. J. Mod. Phys {\bf A7} (1992) 7419\\
                 I. I. Kogan and A. Kovner, Phys. Rev.  {\bf D51} (1995) 1948
\bibitem{P_1}    A.M. Polyakov, Mod. Phys. Lett. {\bf A3} (1988) 325
\bibitem{POL_1}  J. Polchinski, \textit{String Theory}, Cambridge University Press
\bibitem{FMS}    P. Di Francesco, P. Mathieu and D. S\'{e}n\'{e}chal, \textit{Conformal Field Theory}, Springer
\bibitem{R_1}    A. Giveon,M. Porrati and E. Rabinovici, Phys. Rept. {\bf 244} (1994) 77-202
\bibitem{Z_1}    A. Cappelli, C. Itzykson, J.B. Zuber, Nucl. Phys. {\bf B280} (1987) 445-465; Commun. Math. Phys. {\bf 113} (1987) 1


\end{thebibliography}
\end{document}